\font\fiverm=cmr5
\font\fivebf=cmbx5
\font\fivei=cmmi5

\font\fivesy=cmsy5

\font\sevenrm=cmr7
\font\sevenbf=cmbx7
\font\seveni=cmmi7

\font\sevensy=cmsy7

\font\ninerm=cmr9
\font\ninebf=cmbx9

\font\nineit=cmmi9
\font\ninei=cmmi9 
\font\ninesy=cmsy9  
\font\nineex=cmex10
\font\tenrm=cmr10
\font\tenbf=cmbx10
\font\tensl=cmsl10
\font\tenit=cmmi10
\font\teni=cmmi10 
 
\font\tensy=cmsy10
\font\tenex=cmex10
\font\twelverm=cmr12
\font\twelvebf=cmbx12
\font\twelvesl=cmsl12
\font\twelveit=cmmi12
\font\twelvei=cmmi12
\font\twelvesy=cmsy10 scaled\magstep1


%
%
 %
%
 \def\ninepoint{%
   \normalbaselineskip=11pt
   \def\rm{\fam0\ninerm}%
   \def\it{\fam0\nineit}%
   \def\bf{\fam\bffam\ninebf}%
   \def\bi{\fam\bffam\ninebf}%
   \def\rmit{\fam0\ninerm\def\it{\fam0\nineit}}%
   \def\bfit{\fam\bffam\ninebf\def\it{\bi}}%
   \def\bsl{\fam\bffam\ninebsl}
   \textfont0=\ninerm\scriptfont0=\sevenrm\scriptscriptfont0=\fiverm
   \textfont1=\ninei\scriptfont1=\seveni\scriptscriptfont1=\fivei
   \textfont2=\ninesy\scriptfont2=\sevensy\scriptscriptfont2=\fivesy
   \textfont3=\nineex \scriptfont3=\tenex \scriptscriptfont3=\tenex
   \textfont\bffam=\ninebf\scriptfont\bffam=\sevenbf\scriptscriptfont\bffam=
     \fivebf
   \normalbaselines\rm}%
%
 \def\tenpoint{%
   \normalbaselineskip=12pt
   \def\rm{\fam0\tenrm}%
   \def\it{\fam0\tenit}%
   \def\bf{\fam\bffam\tenbf}%
   \def\bi{\fam\bffam\tenbf}%
   \def\rmit{\fam0\tenrm\def\it{\fam0\tenit}}%
   \def\bfit{\fam\bffam\tenbf\def\it{\bi}}%
   \def\bsl{\fam\bffam\tenbsl}
   \textfont0=\tenrm\scriptfont0=\sevenrm\scriptscriptfont0=\fiverm
   \textfont1=\teni\scriptfont1=\seveni\scriptscriptfont1=\fivei
   \textfont2=\tensy\scriptfont2=\sevensy\scriptscriptfont2=\fivesy
   \textfont3=\tenex \scriptfont3=\tenex \scriptscriptfont3=\tenex
     \fivebf
   \textfont\bffam=\tenib\scriptfont\bffam=\sevenib\scriptscriptfont\bffam=
     \fiveib
   \normalbaselines\rm}%
%

%

     
\font\twelverm=cmr10 scaled 1200    \font\twelvei=cmmi10 scaled 1200
\font\twelvesy=cmsy10 scaled 1200   \font\twelveex=cmex10 scaled 1200
\font\twelvebf=cmbx10 scaled 1200   \font\twelvesl=cmsl10 scaled 1200
\font\twelvett=cmtt10 scaled 1200   \font\twelveit=cmti10 scaled 1200
     
\skewchar\twelvei='177   \skewchar\twelvesy='60
     
     
\def\twelvepoint{\normalbaselineskip=12.4pt
  \abovedisplayskip 12.4pt plus 3pt minus 9pt
  \belowdisplayskip 12.4pt plus 3pt minus 9pt
  \abovedisplayshortskip 0pt plus 3pt
  \belowdisplayshortskip 7.2pt plus 3pt minus 4pt
  \smallskipamount=3.6pt plus1.2pt minus1.2pt
  \medskipamount=7.2pt plus2.4pt minus2.4pt
  \bigskipamount=14.4pt plus4.8pt minus4.8pt
  \def\rm{\fam0\twelverm}          \def\it{\fam\itfam\twelveit}%
  \def\sl{\fam\slfam\twelvesl}     \def\bf{\fam\bffam\twelvebf}%
  \def\mit{\fam 1}                 \def\cal{\fam 2}%
  \def\tt{\twelvett}
  \textfont0=\twelverm   \scriptfont0=\tenrm   \scriptscriptfont0=\sevenrm
  \textfont1=\twelvei    \scriptfont1=\teni    \scriptscriptfont1=\seveni
  \textfont2=\twelvesy   \scriptfont2=\tensy   \scriptscriptfont2=\sevensy
  \textfont3=\twelveex   \scriptfont3=\twelveex  \scriptscriptfont3=\twelveex
  \textfont\itfam=\twelveit
  \textfont\slfam=\twelvesl
  \textfont\bffam=\twelvebf \scriptfont\bffam=\tenbf
  \scriptscriptfont\bffam=\sevenbf
  \normalbaselines\rm}
     
     
\def\tenpoint{\normalbaselineskip=12pt
  \abovedisplayskip 12pt plus 3pt minus 9pt
  \belowdisplayskip 12pt plus 3pt minus 9pt
  \abovedisplayshortskip 0pt plus 3pt
  \belowdisplayshortskip 7pt plus 3pt minus 4pt
  \smallskipamount=3pt plus1pt minus1pt
  \medskipamount=6pt plus2pt minus2pt
  \bigskipamount=12pt plus4pt minus4pt
  \def\rm{\fam0\tenrm}          \def\it{\fam\itfam\tenit}%
  \def\sl{\fam\slfam\tensl}     \def\bf{\fam\bffam\tenbf}%
  \def\smc{\tensmc}             \def\mit{\fam 1}%
  \def\cal{\fam 2}%
  \textfont0=\tenrm   \scriptfont0=\sevenrm   \scriptscriptfont0=\fiverm
  \textfont1=\teni    \scriptfont1=\seveni    \scriptscriptfont1=\fivei
  \textfont2=\tensy   \scriptfont2=\sevensy   \scriptscriptfont2=\fivesy
  \textfont3=\tenex   \scriptfont3=\tenex     \scriptscriptfont3=\tenex
  \textfont\itfam=\tenit
  \textfont\slfam=\tensl
  \textfont\bffam=\tenbf \scriptfont\bffam=\sevenbf
  \scriptscriptfont\bffam=\fivebf
  \normalbaselines\rm}
     

{\obeylines\gdef\
{}}
\def\singlespace{\baselineskip=\normalbaselineskip}

\def\doublespace{\baselineskip=\normalbaselineskip \multiply\baselineskip by 2}

\newcount\firstpageno
\firstpageno=2
\footline={\ifnum\pageno<\firstpageno{\hfil}\else{\hfil\twelverm\folio\hfil}\fi}
\let\rawfootnote=\footnote              
\def\footnote#1#2{{\rm\singlespace\parindent=0pt\rawfootnote{#1}{#2}}}

     
\hsize=6.5truein
\hoffset=0truein
\vsize=8.9truein
\voffset=0truein
\parskip=\medskipamount
\twelvepoint            
\doublespace            
\overfullrule=0pt       
     
     
\def\preprintno#1{
 \rightline{\rm #1}}    
     
     
\def\ref#1{Ref. #1}                     

\def\frac#1#2{{\textstyle{#1 \over #2}}}

\def\sla{\raise.15ex\hbox{$/$}\kern-.57em}
\def\leaderfill{\leaders\hbox to 1em{\hss.\hss}\hfill}
\def\twiddle{\lower.9ex\rlap{$\kern-.1em\scriptstyle\sim$}}
\def\bigtwiddle{\lower1.ex\rlap{$\sim$}}
\def\gtwid{\mathrel{\raise.3ex\hbox{$>$\kern-.75em\lower1ex\hbox{$\sim$}}}}
\def\ltwid{\mathrel{\raise.3ex\hbox{$<$\kern-.75em\lower1ex\hbox{$\sim$}}}}
\def\square{\kern1pt\vbox{\hrule height 1.2pt\hbox{\vrule width 1.2pt\hskip 3pt
   \vbox{\vskip 6pt}\hskip 3pt\vrule width 0.6pt}\hrule height 0.6pt}\kern1pt}

\def\tablerule{\tablespace\noalign{\hrule}\tablespace}

\def\m@th{\mathsurround=0pt }
\def\leftrightarrowfill{$\m@th \mathord\leftarrow \mkern-6mu
 \cleaders\hbox{$\mkern-2mu \mathord- \mkern-2mu$}\hfill
 \mkern-6mu \mathord\rightarrow$}
\def\overleftrightarrow#1{\vbox{\ialign{##\crcr
     \leftrightarrowfill\crcr\noalign{\kern-1pt\nointerlineskip}
     $\hfil\displaystyle{#1}\hfil$\crcr}}}

\singlespace
\preprintno{gr-qc/9801028}
\preprintno{CRETE-97-17}
\preprintno{ENS-97/67}
\preprintno{UFIFT-HEP-97-27}
\vskip 2cm
\centerline{\bf PERTURBATIVE QUANTUM GRAVITY AND NEWTON'S LAW}
\centerline{\bf ON A FLAT ROBERTSON-WALKER BACKGROUND}
\vskip 2cm
\centerline{\bf J. Iliopoulos}
\vskip .5cm
\centerline{\it Laboratoire de Physique Th\'eorique,
 Ecole Normale Sup\'erieure}
\centerline{\it 24, rue Lhomond, F-75231 Paris Cedex 05, FRANCE}
\centerline{\rm ilio@physique.ens.fr}
\vskip 1cm
\centerline{\bf T. N. Tomaras , N. C. Tsamis}
\vskip .5cm
\centerline{\it Department of Physics, University of Crete}
\centerline{\it and Research Center of Crete}
\centerline{\it P. O. Box 2208, GR-71003 Heraklion, GREECE}
\centerline{\rm tomaras@physics.uch.gr , tsamis@physics.uch.gr}
\vskip 1cm
\centerline{\bf R. P. Woodard}
\vskip .5cm
\centerline{\it Department of Physics, University of Florida}
\centerline{\it Gainesville, FL 32611, USA}
\centerline{\rm woodard@phys.ufl.edu}
\vskip 2cm
\centerline{ABSTRACT}
\itemitem{}{\tenpoint We derive the Feynman rules for the graviton 
in the presence of a flat Robertson-Walker background and give 
explicit expressions for the propagator in the physically interesting 
cases of inflation, radiation domination, and matter domination. 
The aforementioned background is generated by a scalar field source 
which should be taken to be dynamical. As an elementary application, 
we compute the corrections to the Newtonian gravitational force in 
the present matter dominated era and conclude -- as expected -- 
that they are negligible except for the largest scales.}

\vfill\eject
\doublespace

\centerline{\bf 1. Introduction}

Of the non-trivial spacetime backgrounds, those of particular interest
in cosmology are spatially homogeneous and isotropic. This is a direct
consequence of the cosmological principle which states that there is no 
preferred position in the universe. It can be shown that the most general
background satisfying the above requirements is the Robertson-Walker one
[1]:
$$d{\widehat s}^2 = -dt^2 + a^2(t) \; \Bigl\{ \;
(1 - kr^2)^{-1} \; dr^2 + r^2 d\Omega_2^2 \; \Bigr\}
\;\; . \eqno(1.1)$$
It is characterized by a function $a(t)$ and a constant $k$ which can 
be chosen to take on the values $+1$, $0$, or $-1$. The function $a(t)$
-- known as the scale factor -- is a measure of the ``radius'' of the
universe, while the constant $k$ determines the spatial curvature.

By far the most natural explanation of the observed homogeneity and 
isotropy in the present universe is the assumption of an inflationary
era in the evolution of the universe. During that era, the physical
distance between observers at rest on fixed spatial coordinates increases
superluminally. Accordingly, the observed universe was once so small that 
causal processes could have established an initial thermal equilibrium
across it. This accounts for the fact that the cosmic microwave background
radiation from different regions of the sky is seen to be in thermal
equilibrium to within one part in $10^5$ [2]. Since any decent amount of
inflation
\footnote{*}{\tenpoint By decent we mean at least the $60$ e-foldings 
of inflation needed to explain the observed degree of isotropy.}
redshifts the spatial curvature to insignificance, we can restrict our
study to the $k=0$ geometries:
$$d{\widehat s}^2 = -dt^2 + a^2(t) \; d{\vec x} \cdot d{\vec x}
\;\; . \eqno(1.2)$$
Such geometries do not represent solutions of the Einstein equations
in vacuum except in the case of flat space where the scale factor is
equal to one. In all other cases a non-trivial stress tensor must be
present. Homogeneity and isotropy constrain the stress tensor to have
only diagonal elements consisting of the density $\rho(t)$ and pressure
$p(t)$: $T_{\mu \nu} = [ \; \rho(t) \; p(t) \; p(t) \; p(t) \; ]$. This
form can arise either from a non-zero temperature matter theory or a 
non-trivial scalar field in zero temperature quantum field theory. To 
preserve the isotropy of space, the matter fields must be scalar since
the Lorentz transformation properties of all other fields -- like 
fermions or gauge bosons -- exhibit preferred directions. The 
determination of the scalar field action which supports an arbitrary 
spatially flat Robertson-Walker background is given in Section 2.

It is best to set-up the theory on an arbitrary spatial manifold of 
finite extent -- for instance, $T^3 \times \Re$ -- in which case 
$\Delta x^i \leq {\rm R}$. We shall find it very convenient to utilize 
the conformal set of coordinates:
$$d{\widehat s}^2 = -dt^2 + a^2(t) \; d{\vec x} \cdot d{\vec x}
= \Omega^2(\eta) \; \Bigl( -d\eta^2 + d{\vec x} \cdot d{\vec x} \Bigr)
\;\; . \eqno(1.3a)$$
The relation between the two coordinate systems is given by:
$$dt = \Omega(\eta) \; d\eta 
\qquad ; \qquad
a(t) = \Omega(\eta)
\;\; , \eqno(1.3b)$$
so that a generic power law in co-moving coordinates:
$$a(t) = \Bigl( {t \over t_0} \Bigr)^s
\;\; , \eqno(1.4a)$$
takes the following form in conformal coordinates:
$$\Omega(\eta) = \Bigl( {\eta \over \eta_0} \Bigr)^{\frac{s}{1-s}}
\;\; , \eqno(1.4b)$$
with $\eta_0 = t_0 \; (1-s)^{-1}$. The physically most distinguishable 
cases consist of the {\it inflating} universe for $s = +\infty$ , the 
{\it radiation dominated} universe for $s = \frac12$ , the {\it matter 
dominated} universe for $s = \frac23$ , and the {\it flat} universe for 
$s = 0$. We shall also assume throughout that $\Delta x \ll {\rm R}$
and $\Delta \eta \ll {\rm R}$ since then the integral approximation to 
mode sums which appear because we work on $T^3$ is excellent.
\footnote{*}{\tenpoint In conformal coordinates light reaches the 
distance $\Delta x = {\rm R}$ in time $\Delta \eta = {\rm R}$.} 

In order to study gravitational effects not associated with the 
smallest of scales in the class of backgrounds (1.2), we must develop 
the appropriate perturbative tools -- this is accomplished in Section 3.
As an elementary application of the perturbative results, we calculate 
in Section 4 the response of the gravitational field due to a point 
source in a matter dominated universe. The resulting corrections to the 
Newtonian long-range gravitational force are found to be negligible as
expected. Our conclusions comprise Section 5.

\vskip 1cm
\centerline{\bf 2. Determination of the source background}

The dynamical system under consideration has a background action which 
consists of two parts:
$${\cal \widehat S} = 
{\cal \widehat S}_{\rm g} + {\cal \widehat S}_{\rm m}
\;\; . \eqno(2.1)$$
The gravitational action ${\cal \widehat S}_{\rm g}$ is obtained from the 
Einstein Lagrangian:
$${\cal \widehat L}_{\rm g} =
{1 \over 16\pi G} \; R \; \sqrt {-g}
\;\; , \eqno(2.2a)$$
while the matter action ${\cal \widehat S}_{\rm m}$ is based on a generic 
scalar field Lagrangian:
$${\cal \widehat L}_{\rm m} =
- \frac12 \; \partial_{\mu} \varphi \; \partial_{\nu} \varphi \;
g^{\mu \nu} \sqrt {-g} \;
- \; V(\varphi) \; \sqrt {-g}
\;\; . \eqno(2.2b)$$
We obtain the following equations of motion for the gravitational field:
$$G^{\mu \nu} = (8\pi G) \; T^{\mu \nu}
\;\; , \eqno(2.3a)$$
and for the scalar field:
$$\partial_{\mu} \Bigl( 
\sqrt {-g} \; g^{\mu \nu} \; \partial_{\nu} \varphi \Bigr)
= {\partial V(\varphi) \over \partial \varphi} \; \sqrt {-g}
\;\; . \eqno(2.3b)$$
The Einstein tensor $G^{\mu \nu}$ is determined by the geometry (1.2):
\footnote{*}{\tenpoint  Our metric has spacelike signature and
$R^{\mu}_{~\nu\rho\sigma} = \Gamma^{\mu}_{~\sigma\nu , \rho} + ... \;$ 
Greek indices run from $0$ to $3$, while Latin indices run from $1$ to $3$. 
Dots as superscripts denote differentiation with respect to the co-moving 
time $t$. We are also using units where $c=\hbar=1$.} 
$$G^{00} = 3 \; \Biggl( {{\dot a} \over a} \Biggr)^2
\qquad ; \qquad
G^{ij} = -g^{ij} \; \Biggl( \;
2 { {\ddot a} \over a } +
{ {\dot a}^2 \over a^2 } \; \Biggr)
\;\; , \eqno(2.4)$$
and the covariantly conserved stress tensor $T^{\mu \nu}$ by the matter
sector of the theory:
$$T^{\mu \nu} \equiv
{2 \over \sqrt {-g} } \; 
{\delta {\cal S}_{\rm m} \over \delta g_{\mu \nu} } =
\Bigl( \; g^{\mu \alpha} \; g^{\nu \beta} -
\frac12 g^{\mu \nu} \; g^{\alpha \beta} \; \Bigr) \;
\partial_{\alpha} \varphi \; \partial_{\beta} \varphi \; - \;
g^{\mu \nu} \; V(\varphi) 
\;\; , \eqno(2.5a)$$
$$T_{~~~ ; \nu}^{\mu \nu} = 0
\;\; . \eqno(2.5b)$$
The resulting density $\rho(t)$ and pressure $p(t)$ take the form:
$$T^{00} \equiv \rho =
\frac12 {\dot {\widehat \phi}}^{\;2} + V({\widehat \phi})
\;\; , \eqno(2.6a)$$
$$T^{ij} \equiv p \; g^{ij} =
g^{ij} \; \Bigl( \; \frac12 {\dot {\widehat \phi}}^{\;2} - 
V({\widehat \phi}) \; \Bigr) \;\; . \eqno(2.6b)$$
In order not to disturb the spatial homogeneity and isotropy of our
background geometries, we have taken the scalar field to be a function
of time only: $\varphi = {\widehat \phi}(t)$.

Using (2.4) and (2.6), the gravitational equations (2.3a) become:
$${\dot {\widehat \phi}}^{\;2} = {1 \over 4\pi G} \;
\Biggl( \; - { {\ddot a} \over a } +
{ {\dot a}^2 \over a^2 } \; \Biggr)
\;\; , \eqno(2.7a)$$
$$V({\widehat \phi}) = {1 \over 8\pi G} \;
\Biggl( \; { {\ddot a} \over a } +
2 { {\dot a}^2 \over a^2 } \; \Biggr)
\;\; . \eqno(2.7b)$$
By satisfying equations (2.7), we determine the source parameters
${\dot {\widehat \phi}}(t)$ and $V({\widehat \phi})$ -- 
or, equivalently, $\rho(t)$ and 
$p(t)$ -- which support the flat Robertson-Walker background (1.2). 
In conformal coordinates the above equations take the form:
\footnote{*}{\tenpoint Notice that in the case of de Sitter spacetime,
$s \rightarrow +\infty$, equations (2.7-8) have the proper limit:
${\dot {\widehat \phi}}^{\;2} \rightarrow 0$ and $V({\widehat \phi}) 
\rightarrow (8\pi G) \; 
{\Lambda}^{-1}$, where $\Lambda = 3 \eta_0^{-2}$.}
$${\dot {\widehat \phi}}^{\;2} = {1 \over 4\pi G} \;
\Biggl( \; - {\Omega'' \over \Omega^3} +
{2 {\Omega'}^2  \over \Omega^4} \; \Biggr)
\;\; , \eqno(2.8a)$$
$$V({\widehat \phi}) = {1 \over 8\pi G} \;
\Biggl( \; {\Omega'' \over \Omega^3 } +
{ {\Omega'}^2 \over \Omega^4} \; \Biggr)
\;\; , \eqno(2.8b)$$
where the primes over $\Omega$ indicate differentiation with respect to 
the conformal time $\eta$. It remains to be shown that the above choice
for the source parameters is consistent with the scalar equation of
motion (2.3b):
$${\ddot {\widehat \phi}} + {3 {\dot a} \over a} \; {\dot {\widehat \phi}} +
{\partial V({\widehat \phi}) \over \partial {\widehat \phi}} = 0
\;\; . \eqno(2.9)$$
This is a direct consequence of the conservation equation (2.5b):
$${\dot \rho} = - {3 {\dot a} \over a} \; (\rho + p)
\;\; . \eqno(2.10)$$
By substituting (2.6) in (2.10) we obtain:
$${d \over dt} \Bigl( \; \frac12 {\dot {\widehat \phi}}^{\;2} + 
V({\widehat \phi}) \; \Bigr) 
= - {3 {\dot a} \over a} \; {\dot {\widehat\phi}}^{\;2} 
\;\; , \eqno(2.11)$$
which is identical to (2.9).

\vskip 1cm
\centerline{\bf 3. The Feynman rules}

{\it 3.1 The Lagrangian.}

The quantum fields are the graviton $h_{\mu \nu}(x)$ and the scalar
$\phi(x)$:
$$ g_{\mu \nu} \equiv {\widehat g}_{\mu \nu} + \kappa h_{\mu \nu}
\;\; , \eqno(3.1a)$$
$$\varphi \equiv {\widehat \phi} + \phi
\; \; . \eqno(3.1b)$$
It turns out that -- just like the case of de Sitter spacetime [3] --
it is most convenient to organize perturbation theory in terms of the
``pseudo-graviton'' field, $\psi_{\mu \nu}(x)$, obtained by conformally
re-scaling the metric:
$$g_{\mu \nu} \equiv 
\Omega^2 \; {\widetilde g}_{\mu \nu} \equiv 
\Omega^2 \; \Bigl( \eta_{\mu \nu} + \kappa \psi_{\mu \nu} \Bigr)
\;\; . \eqno(3.2)$$
As usual, pseudo-graviton indices are raised and lowered with the 
Lorentz metric and $\kappa^2 \equiv 16\pi G$. The total Lagrangian is:
$${\cal L} = {1 \over \kappa^2} \; R \; \sqrt {-g} \; - \; 
\frac12 \; \partial_{\mu} \varphi \; \partial_{\nu} \varphi \;
g^{\mu \nu} \sqrt {-g} \; - \;
V(\varphi) \; \sqrt {-g}
\;\; . \eqno(3.3)$$
By substituting (3.1b) in (3.3) we get:
$$\eqalignno{
{\cal L} = {1 \over \kappa^2} \; R \; \sqrt {-g} 
\; - \; \frac12 {\widehat \phi}^{\;'2} \; g^{00} \sqrt {-g}
\; - \; \partial_{\mu} {\widehat \phi} \; \partial_{\nu} \phi \;
g^{\mu \nu} \sqrt {-g}
&\; - \; \frac12 \partial_{\mu} \phi \; \partial_{\nu} \phi \;
g^{\mu \nu} \sqrt {-g} \cr
-& \sum_{n=0}^{+\infty} \; {1 \over n!} \; 
{\partial^n V \over \partial \varphi^n}({\widehat \phi}) \;\; 
\phi^n \sqrt {-g}
\;\; , \qquad &(3.4) }$$
and we can proceed to organize ${\cal L}$ according to the number of
fields $\phi$ present.

The pure metric part of ${\cal L}$:
$${\cal L}_{(0)} = {1 \over \kappa^2} \; R \; \sqrt {-g}
\; - \; \frac12 {\widehat \phi}^{\;'2} \; g^{00} \sqrt {-g}
\; - \; V({\widehat \phi}) \; \sqrt {-g}
\;\; , \eqno(3.5)$$
in terms of the re-scaled metric is:
$${\cal L}_{(0)} = {1 \over \kappa^2} \; \Biggl\{ \;
\Omega^2 \; {\widetilde R} +
2 \Bigl( -1 + {\widetilde g}^{00} \Bigr)
\Bigl( \Omega \Omega'' + {\Omega'}^2 \Bigr) \;
\Biggr\} \; \sqrt {- \widetilde g}
\;\; , \eqno(3.6)$$
where we used (2.8) and ignored surface terms. After performing many 
partial integrations, (3.6) can be cast in a form identical to that 
of de Sitter spacetime [3]:
$$\eqalignno{
{\cal L}_{(0)} = \sqrt {- \widetilde g} \;
{\widetilde g}^{\alpha \beta} \; {\widetilde g}^{\rho \sigma} \;
{\widetilde g}^{\mu \nu} \Bigl[ 
\frac12 \psi_{\alpha \rho , \mu} \; \psi_{\nu \sigma , \beta} \;
-\frac12 \psi_{\alpha \beta, \rho} \; & \psi_{\sigma \mu , \nu} \;
+\frac14 \psi_{\alpha \beta , \rho} \; \psi_{\mu \nu , \sigma} \;
-\frac14 \psi_{\alpha \rho , \mu} \; \psi_{\beta \sigma , \nu} 
\Bigr] \Omega^2 \cr
-& \frac12 \sqrt {- \widetilde g} \;
{\widetilde g}^{\rho \sigma} \; {\widetilde g}^{\mu \nu} \;
\psi_{\rho \sigma , \mu} \; \psi_{\nu}^{~\alpha} \; 
(\Omega^2)_{, \alpha} 
\;\; . &(3.7) }$$
All the self-interactions of the graviton in the presence of an arbitrary 
scale factor $\Omega(\eta)$ can be obtained by expanding expression (3.7). 
The quadratic part of ${\cal L}_{(0)}$ is:
$$\eqalignno{
{\cal L}_{(0)}^{(2)} = \Bigl[ \;
\frac12 \psi^{\alpha \rho , \mu} \; \psi_{\mu \rho , \alpha} \;
-\frac12 \psi_{, \alpha} \; \psi^{\alpha \rho}_{~~, \rho} \;
+\frac14 \psi^{, \alpha} \; \psi_{, \alpha} \;
-\frac14 \psi^{\alpha \rho , \mu} \;& \psi_{\alpha \rho , \mu} \;
\Bigr] \; \Omega^2 \cr
+& \psi^{, \mu} \; \psi_{\mu 0} \; \Omega \; \Omega'
\;\; , &(3.8) }$$
where $\psi \equiv \psi^{\mu}_{~\mu}$.

The part of ${\cal L}$ linear in $\phi$ is simpler:
$$\eqalignno{
{\cal L}_{(1)} &=
- {\widehat \phi}^{\;'} \; \partial_{\nu} \phi \;
g^{0 \nu} \sqrt {-g} \; - \; 
{\partial V \over \partial \varphi}({\widehat \phi}) \; 
\phi \; \sqrt {-g}
\;\; , &(3.9a) \cr
&= {1 \over \kappa} \Bigl[ \;
- \xi \; \partial_{\mu} \phi \; 
{\widetilde g}^{\mu 0} + \xi' \; \phi \; \Bigr] 
\sqrt {- \widetilde g}
\;\; . &(3.9b) }$$ 
In the last step we used the scalar background equation of motion
(2.9) to derive the identity:
$${\partial V \over \partial \varphi}({\widehat \phi}) =
-{1 \over \kappa \Omega^4} \; \xi'
\;\; , \eqno(3.10a)$$
where we define $\xi(\eta)$ as:
$$\xi \equiv \kappa \Omega^2 \; {\widehat \phi}^{\;'}
\;\; . \eqno(3.10b)$$
The following quadratic part emerges:
$${\cal L}_{(1)}^{(2)} =
\xi \; \partial_{\mu} \phi \; \psi^{\mu 0} +
\frac12 \; (\xi \phi)' \; \psi
\;\; . \eqno(3.11)$$

We shall also need the part of ${\cal L}$ which is quadratic in
$\phi$:
$$\eqalignno{
{\cal L}_{(2)} &=
- \frac12 \partial_{\mu} \phi \; \partial_{\nu} \phi \;
g^{\mu \nu} \sqrt {-g} \; - \; 
\frac12 {\partial^2 V \over \partial \varphi^2}({\widehat \phi}) \; 
\phi^2 \; \sqrt {-g}
\;\; , &(3.12a) \cr
&= - \frac12 \Omega^2 \; 
\partial_{\mu} \phi \; \partial_{\nu} \phi \; 
{\widetilde g}^{\mu \nu} \sqrt {- \widetilde g} \; + \; 
\frac12 \Omega^6 \; \xi^{-1} \; (\; \Omega^{-4} \xi' \;)' \; 
\phi^2 \; \sqrt {- \widetilde g}
\;\; , &(3.12b) }$$ 
and the associated piece which is quadratic in the quantum fields:
$${\cal L}_{(2)}^{(2)} =
- \frac12 \Omega^2 \; 
\partial_{\mu} \phi \; \partial^{\mu} \phi \; 
\; + \; 
\frac12 \Omega^6 \; \xi^{-1} \; (\; \Omega^{-4} \xi' \;)' \; 
\phi^2 
\;\; . \eqno(3.13)$$ 

In order to properly quantize the theory and calculate the various 
propagators, we must fix the gauge. We shall accomplish this by 
adding the following term to the Lagrangian:
$${\cal L}_{\rm GF} =
- \frac12 \eta_{\mu \nu} \; F^{\mu} \; F^{\nu}
\;\; , \eqno(3.14a)$$
where:
$$F_{\mu} = 
\Omega \; \psi_{\mu ~ , \nu}^{~ \nu}
- \frac12 \Omega \; \psi_{, \mu}
- 2 \Omega' \; \psi_{\mu 0}
+ \Omega^{-1} \; \eta_{\mu 0} \; \xi \; \phi
\;\; . \eqno(3.14b)$$
The resulting gauge-fixed quadratic Lagrangian is:
$$\eqalignno{
{\cal L}_{\rm GF}^{(2)} =&
{\cal L}_{(0)}^{(2)} + {\cal L}_{(1)}^{(2)}
+ {\cal L}_{(2)}^{(2)} + {\cal L}_{\rm GF} \cr
=& - \frac18 \psi \; \Omega
\Biggl[ \; \partial^2 + {\Omega'' \over \Omega} \; \Biggr]
\Omega \; \psi 
+ \frac14 \psi^{\mu \nu} \; \Omega
\Biggl[ \; \partial^2 + {\Omega'' \over \Omega} \; \Biggr]
\Omega \; \psi_{\mu \nu} \cr
&+ \psi^{\mu 0} \; \Omega
\Biggl[ \; -{\Omega'' \over \Omega} +
{{\Omega'}^2 \over \Omega^2} \; \Biggr]
\Omega \; \psi_{\mu 0}
+ \psi_{00} \; \Omega
\Biggl[ \; -\Omega^{-2} \; \xi' + 
2 {\Omega' \over \Omega^3} \; \xi \; \Biggr]
\Omega \; \phi \cr
&+ \frac12 \phi \; \Omega
\Biggl[ \; \partial^2 + {\Omega'' \over \Omega} + 
\Omega^{-4} \; \xi^2 + 
\Omega^4 \; \xi^{-1} \; (\; \Omega^{-4} \xi' \;)' \; \Biggr]
\Omega \; \phi
\;\; . &(3.15) }$$
In terms of a kinetic operator $D_{\mu \nu}^{~~ \rho \sigma}$ 
we get:
$$\eqalignno{
{\cal L}_{\rm GF}^{(2)} \equiv&
+ \frac12 \psi^{\mu \nu} \; D_{\mu \nu}^{~~ \rho \sigma} \;
\psi_{\rho \sigma}
+ \psi_{00} \; \Omega
\Biggl[ \; -\Omega^{-2} \; \xi' + 
2 {\Omega' \over \Omega^3} \; \xi \; \Biggr]
\Omega \; \phi \cr
&+ \frac12 \phi \; \Omega
\Biggl[ \; \partial^2 + {\Omega'' \over \Omega} + 
\Omega^{-4} \; \xi^2 + 
\Omega^4 \; \xi^{-1} \; (\; \Omega^{-4} \xi' \;)' \; \Biggr]
\Omega \; \phi
\;\; . &(3.16) }$$
where we have:
$$\eqalignno{
D_{\mu \nu}^{~~\rho \sigma} =
\Bigl[ \; \frac12 {\overline \delta}_{\mu}^{~(\rho} \;
{\overline \delta}_{\nu}^{~\sigma)} -
\frac14 \eta_{\mu \nu} \; \eta^{\rho \sigma} -
\frac12 \delta_{\mu}^{~0} \; \delta_{\nu}^{~0} \;
\delta_0^{~\rho} \; \delta_0^{~\sigma} \; \Bigr] \;
&{\rm D}_A \cr
+ \delta_{(\mu}^{~~0} \; {\overline \delta}_{\nu)}^{~~(\rho} \;
\delta_0^{~\sigma)} \; {\rm D}_B
+ \delta_{\mu}^{~0} \; \delta_{\nu}^{~0} \;
\delta_0^{~\rho} \; \delta_0^{~\sigma} \; &D_B
\;\; . \qquad &(3.17) }$$
Parenthesized indices are symmetrized and a bar above a Lorentz metric
or a Kronecker delta symbol means that the zero (i.e., $\eta$) component
is projected out:
$${\overline \eta}_{\mu \nu} \equiv 
\eta_{\mu \nu} + \delta_{\mu}^{~0} \; \delta_{\nu}^{~0}
\qquad ; \qquad
{\overline \delta}_{\mu}^{~\nu} \equiv
\delta_{\mu}^{~\nu} - \delta_{\mu}^{~0} \; \delta_0^{~\nu}
\;\; . \eqno(3.18)$$
The quadratic operators defined in (3.17) are given by:
$$\eqalignno{
{\rm D}_A &\equiv
\Omega \Biggl[ \;
\partial^2 + {\Omega'' \over \Omega}
\; \Biggr] \Omega 
\;\; , &(3.19a) \cr
{\rm D}_B &\equiv
\Omega \Biggl[ \;
\partial^2 - {\Omega'' \over \Omega} + 2 {{\Omega'}^2 \over \Omega^2}
\; \Biggr] \Omega
\;\; . &(3.19b) }$$
Notice that ${\rm D}_A$ is the kinetic operator for a massless, minimally
coupled scalar in the presence of the gravitational background (1.3).

{\it 3.2 The Diagonal Variables.}

The quadratic Lagrangian (3.16) should be brought into diagonal form. 
There is mixing in the first term of (3.16) between $\psi_{ii}$ and 
$\psi_{00}$:
$$\eqalignno{
\frac12 \psi^{\mu \nu} \; D_{\mu \nu}^{~~ \rho \sigma} \;
\psi_{\rho \sigma} 
= \frac12 \Bigl\{ \; 
\frac12 \psi_{ij} \; {\rm D}_A \; \psi_{ij}
&- \frac14 ( \psi_{ii} - \psi_{00} ) \; {\rm D}_A \; 
( \psi_{jj} - \psi_{00} )
- \frac12 \psi_{00} \; {\rm D}_A \; \psi_{00} \cr
&- \psi_{0i} \; {\rm D}_B \; \psi_{0i}
+ \psi_{00} \; D_B \; \psi_{00}
\; \Bigr\}
\;\; , &(3.20) }$$
which is removed by transforming to the field variable 
$\zeta_{\mu \nu}$:
$$\eqalignno{
\frac12 \psi^{\mu \nu} \; D_{\mu \nu}^{~~ \rho \sigma} \;
\psi_{\rho \sigma} 
&= \frac12 \Bigl\{ \;
\frac12 \zeta_{ij} \; {\rm D}_A \; \zeta_{ij}
- \frac14 \zeta_{ii} \; {\rm D}_A \; \zeta_{jj}
- \zeta_{0i} \; {\rm D}_B \; \zeta_{0i}
+ \zeta_{00} \; D_B \; \zeta_{00} 
\; \Bigr\} \cr
&= \frac12 \Bigl\{ \;
\zeta_{ij} \; {\rm D}_A \; \zeta_{rs} \;
\Bigl[ \frac12 \delta_{i ( r} \; \delta_{s ) j} -
\frac14 \delta_{ij} \; \delta_{rs} \Bigr]
+ \zeta_{0i} \; {\rm D}_B \; \zeta_{0r} \; 
\Bigl[ - \delta_{ir} \Bigr] \cr
&\qquad\qquad\qquad\qquad\qquad\qquad
+ \zeta_{00} \; D_B \; \zeta_{00} 
\; \Bigr\} 
\;\; , &(3.21) }$$
where we defined:
$$\psi_{ij} \equiv \zeta_{ij} + \delta_{ij} \; \zeta_{00}
\qquad \; \qquad
\psi_{0i} \equiv \zeta_{0i}
\qquad \; \qquad
\psi_{00} \equiv \zeta_{00}
\;\; . \eqno(3.22)$$

There is also mixing between $\zeta_{00}$ and $\phi$. The part 
of the quadratic Lagrangian involved is:
$$\eqalignno{
{\cal L}_{\rm mixing} =&
\; \frac12 \zeta_{00} \; D_B \; \zeta_{00} 
+ \zeta_{00} \; \Omega
\Biggl[ \; -\Omega^{-2} \; \xi' + 
2 {\Omega' \over \Omega^3} \; \xi \; \Biggr]
\Omega \; \phi \cr
&+ \frac12 \phi \; \Omega
\Biggl[ \; \partial^2 + {\Omega'' \over \Omega} + 
\Omega^{-4} \; \xi^2 + 
\Omega^4 \; \xi^{-1} \; (\; \Omega^{-4} \xi' \;)' \; \Biggr]
\Omega \; \phi
\;\; . &(3.23) }$$
We can always diagonalize this system but, for most scale factors
$\Omega(\eta)$, the resulting kinetic operators will be non-local. In this
case the mode functions obey fourth order differential equations. However,
for a class of scale factors which includes the generic power law (1.4),
the system can be diagonalized without sacrificing locality or the second
order character of the mode equations.\footnote{*}{\tenpoint The condition 
that $\Omega(\eta)$ must obey is:
$- \Omega \; {d \over d\eta} \Bigl( \Omega^{-2} \; \Omega' \Bigr) =
{c_1 \over c_2} \; \tan^2 \Bigl( {\sqrt {c_1 c_2}} \; \eta +
{\sqrt {c_1 \over c_2}} \; \theta \Bigr)$, where $c_1$,$c_2$ are constants 
and $\theta$ is an angle.}
In this case, ${\cal L}_{\rm mixing}$ takes the form:
$$\eqalignno{
{\cal L}_{\rm mixing}^s =
\; \frac12 \zeta_{00} \; \Omega &\Bigl[ \; 
\partial^2 + \frac{s}{(1-s)^2} \; {1 \over \eta^2}
\; \Bigr] \Omega \; \zeta_{00} 
+ \zeta_{00} \; \Omega \;
\frac{2{\sqrt s}}{1-s} \; {1 \over \eta^2} 
\; \Omega \; \phi \cr
+ \frac12 \phi \; \Omega &\Bigl[ \;
\partial^2 + \frac{2s^2 -3s + 2}{(1-s)^2} \; {1 \over \eta^2}
\; \Bigr] \Omega \; \phi
\;\; . &(3.24) }$$
Let us call the diagonal variables $\chi$ and $\upsilon$:
$$\eqalignno{
\zeta_{00} &= \cos \theta \; \chi - \sin \theta \; \upsilon 
\;\; , &(3.25a) \cr
\phi &= \sin \theta \; \chi + \cos \theta \; \upsilon
\;\; , &(3.25b) }$$ 
and demand that the part of ${\cal L}_{\rm mixing}^s$ involving
$\chi \upsilon$ vanish:
$${\cal L}_{\chi \upsilon}^s =
{1 \over \eta^2} \; \chi \; \Omega \; \Bigl\{ \; 
\sin (2 \theta) + \frac{2\sqrt s}{1-s} \; \cos (2\theta) 
\; \Bigr\} \; \Omega \; \upsilon = 0 
\;\; . \eqno(3.26)$$
Of the two available solutions, it is sufficient to consider one of them:
$$\cos^2 \theta = \frac1{1+s} 
\qquad ; \qquad
\sin^2 \theta = \frac{s}{1+s}
\qquad ; \qquad
\sin \theta \; \cos \theta = - \frac{\sqrt s}{1+s}
\;\; , \eqno(3.27)$$
since the other merely interchanges the role of $\chi$ and $v$. The result 
is the following diagonal form:
$${\cal L}_{\rm mixing}^s =
\frac12 \chi \; \Omega \Bigl[ \;
\partial^2 + \frac{2s^2 -s}{(1-s)^2} \; {1 \over \eta^2}
\; \Bigr] \Omega \; \chi
+ \frac12 \upsilon \; \Omega \Bigl[ \;
\partial^2 + \frac{2-s}{(1-s)^2} \; {1 \over \eta^2}
\; \Bigr] \Omega \; \upsilon
\;\; . \eqno(3.28)$$

The complete diagonal quadratic Lagrangian is:
$$\eqalignno{
{\cal L}_{\rm GF}^{(2)} =&
\; \frac12 \zeta_{ij} \; {\rm D}_A^s \; \zeta_{rs} \;
\Bigl[ \frac12 \delta_{i ( r} \; \delta_{s ) j} -
\frac14 \delta_{ij} \; \delta_{rs} \Bigr]
+ \frac12 \zeta_{0i} \; {\rm D}_B^s \; \zeta_{0r} \; 
\Bigl[ - \delta_{ir} \Bigr] \cr
&+ \frac12 \upsilon \; {\rm D}_C^s \; \upsilon
+ \frac12 \chi \; {\rm D}_A^s \; \chi 
\;\; , &(3.29) }$$ 
where:
$$\eqalignno{
{\rm D}_A^s &= 
\Omega \Bigl[ \; \partial^2 + 
\frac{2s^2 -s}{(1-s)^2} \; {1 \over \eta^2}
\; \Bigr] \Omega  
\;\; , &(3.30a) \cr
{\rm D}_B^s &= 
\Omega \Bigl[ \; \partial^2 + 
\frac{s}{(1-s)^2} \; {1 \over \eta^2}
\; \Bigr] \Omega
\;\; , &(3.30b) \cr
{\rm D}_C^s &= 
\Omega \Bigl[ \; \partial^2 + 
\frac{2-s}{(1-s)^2} \; {1 \over \eta^2}
\; \Bigr] \Omega
\;\; . &(3.30c) }$$ 
It becomes transparent that the ``$A$'' graviton modes are six 
and have purely spatial polarizations, the ``$B$'' graviton modes
are three and have mixed polarizations, and the ``$C$'' graviton
mode is one with purely temporal polarization. Clearly, the two
physical graviton polarizations are transverse traceless ``$A$''
modes.

{\it 3.3 The Propagators.}
 
Manifest spatial translation invariance allows us to write the 
general linearized solution as the following superposition:
$$\eqalignno{
\psi_{\mu \nu}(\eta, {\vec x}) &=
\sum_{\lambda} \int {d^3k \over (2\pi)^3} \;
\exp[i {\vec k} \cdot {\vec x}] \;
\Bigl\{ \;
\Psi(\eta, k, \lambda) \; a_{\mu \nu}({\vec k}, \lambda) +
\Psi^{\star}(\eta, k, \lambda) \; 
a_{\mu \nu}^{\dagger}({\vec k}, \lambda) 
\; \Bigr\} 
\;\; , \cr
\phi(\eta, {\vec x}) &=
\int {d^3k \over (2\pi)^3} \;
\exp[i {\vec k} \cdot {\vec x}] \;
\Bigl\{ \;
\Psi(\eta, k) \; a({\vec k}) +
\Psi^{\star}(\eta, k) \; a^{\dagger}({\vec k}) 
\; \Bigr\} 
\;\; , &(3.31b) }$$
where $k = \Vert {\vec k} \Vert$ and the index $\lambda$ spans all 
available polarizations. The free vacuum of the theory is defined 
as the normalizable state that obeys:
\footnote{*}{\tenpoint Besides the $a_{\mu \nu}({\vec k}, \lambda)$, 
the analogous ghost operators annihilate $\vert 0 \rangle$ as well.} 
$$a_{\mu \nu}({\vec k}, \lambda) \; \Bigr\vert 0 \Bigr\rangle = 
a({\vec k}) \; \Bigr\vert 0 \Bigr\rangle = 0
\;\; . \eqno(3.32)$$
The mode functions are canonically normalized:
$$\Psi(\eta, k, \lambda) \; \Psi^{\star'}(\eta, k, \lambda)
- \Psi^{\star}(\eta, k, \lambda) \; \Psi'(\eta, k, \lambda)
= i\Omega^{-2} \qquad \forall \lambda 
\;\; , \eqno(3.33)$$
and are annihilated upon acting the appropriate quadratic 
operators. In our case:
$${\rm D}_I^s \; \Psi(\eta, k, I) = 0 
\qquad , \qquad I=A,B,C
\;\; . \eqno(3.34)$$
The pseudo-graviton and scalar propagators are defined by:
$$\eqalignno{
i\Bigl[ {_{\mu \nu}}\Delta_{\rho \sigma} \Bigr](x;x') &\equiv
\Bigl\langle 0 \Bigl\vert \; T \Big\{ 
\psi_{\mu \nu}(x) \; \psi_{\rho \sigma}(x') 
\Bigr\} \; \Bigr\vert 0 \Bigr\rangle
\;\; , &(3.35a) \cr
i\Delta (x;x') &\equiv
\Bigl\langle 0 \Bigl\vert \; T \Big\{ 
\phi(x) \; \phi(x') 
\Bigr\} \; \Bigr\vert 0 \Bigr\rangle
\;\; , &(3.35b) }$$ 
but we must consider the mixed propagator as well:
$$i\Bigl[ {_{\mu \nu}}\Delta \Bigr](x;x') \equiv 
\Bigl\langle 0 \Bigl\vert \; T \Big\{ 
\psi_{\mu \nu}(x) \; \phi(x') 
\Bigr\} \; \Bigr\vert 0 \Bigr\rangle 
\;\; . \eqno(3.35c)$$

Naturally, it is the propagators of the diagonal variables
that we shall evaluate at first. From (3.29) we conclude that:
$$\eqalignno{
\Bigl\langle 0 \Bigl\vert \; T \Big\{ 
\zeta_{ij}(x) \; \zeta_{rs}(x') 
\Bigr\} \; \Bigr\vert 0 \Bigr\rangle &=
i\Delta_A^s(x;x') \;\;
2\Bigl[ \delta_{i(r} \; \delta_{s)j} -
\delta_{ij} \; \delta_{rs} \Bigl]
\;\; , &(3.36a) \cr
\Bigl\langle 0 \Bigl\vert \; T \Big\{ 
\zeta_{0i}(x) \; \zeta_{0r}(x') 
\Bigr\} \; \Bigr\vert 0 \Bigr\rangle &=
i\Delta_B^s(x;x') \; \Bigl[ -\delta_{ir} \Bigl]
\;\; , &(3.36b) \cr
\Bigl\langle 0 \Bigl\vert \; T \Big\{ 
\upsilon(x) \; \upsilon(x') 
\Bigr\} \; \Bigr\vert 0 \Bigr\rangle &=
i\Delta_C^s(x;x')
\;\; , &(3.36c) \cr
\Bigl\langle 0 \Bigl\vert \; T \Big\{ 
\chi(x) \; \chi(x') 
\Bigr\} \; \Bigr\vert 0 \Bigr\rangle &=
i\Delta_A^s(x;x')
\;\; , &(3.36d) }$$
where the propagators $i\Delta_I^s(x;x')$ satisfy:
$${\rm D}_I^s \;\; i\Delta_I^s(x;x') =
i\delta^{(4)}(x-x')
\qquad , \qquad I=A,B,C
\;\; . \eqno(3.37)$$
We transform back to the original fields with relations (3.22)
and (3.25):
$$\eqalignno{
\Bigl\langle 0 \Bigl\vert \; T \Big\{ 
\psi_{ij}(x) \; \psi_{rs}(x') 
\Bigr\} \; \Bigr\vert 0 \Bigr\rangle =& \;
i\Delta_A^s(x;x') \;
2\Bigl[ \delta_{i(r} \; \delta_{s)j} -
\delta_{ij} \; \delta_{rs} \Bigl] 
+i\Delta_C^s(x;x') \; \frac{s}{1+s} \;
\delta_{ij} \; \delta_{rs} \cr
&+i\Delta_A^s(x;x') \; \frac{1}{1+s} \;
\delta_{ij} \; \delta_{rs}
\;\; , &(3.38a) \cr 
\Bigl\langle 0 \Bigl\vert \; T \Big\{ 
\psi_{0i}(x) \; \psi_{0r}(x') 
\Bigr\} \; \Bigr\vert 0 \Bigr\rangle =& \;
i\Delta_B^s(x;x') \; \Bigl[ -\delta_{ir} \Bigl]
\;\; , &(3.38b) \cr
\Bigl\langle 0 \Bigl\vert \; T \Big\{ 
\psi_{00}(x) \; \psi_{00}(x') 
\Bigr\} \; \Bigr\vert 0 \Bigr\rangle =& \;
i\Delta_A^s(x;x') \; \frac{1}{1+s}
+ i\Delta_C^s(x;x') \; \frac{s}{1+s}
\;\; , &(3.38c) \cr
\Bigl\langle 0 \Bigl\vert \; T \Big\{ 
\psi_{00}(x) \; \psi_{rs}(x') 
\Bigr\} \; \Bigr\vert 0 \Bigr\rangle =& \;
i\Delta_A^s(x;x') \; \frac{1}{1+s} \; \delta_{rs}
+ i\Delta_C^s(x;x') \; \frac{s}{1+s} \; \delta_{rs}
\;\; , &(3.38d) \cr
\Bigl\langle 0 \Bigl\vert \; T \Big\{ 
\psi_{ij}(x) \; \phi(x') 
\Bigr\} \; \Bigr\vert 0 \Bigr\rangle =& \;
- i\Delta_A^s(x;x') \; \frac{\sqrt s}{1+s} \; \delta_{ij}
+ i\Delta_C^s(x;x') \; \frac{\sqrt s}{1+s} \; \delta_{ij}
\;\; , &(3.38e) \cr
\Bigl\langle 0 \Bigl\vert \; T \Big\{ 
\psi_{00}(x) \; \phi(x') 
\Bigr\} \; \Bigr\vert 0 \Bigr\rangle =& \;
- i\Delta_A^s(x;x') \; \frac{\sqrt s}{1+s}  
+ i\Delta_C^s(x;x') \; \frac{\sqrt s}{1+s} 
\;\; , &(3.38f) \cr
\Bigl\langle 0 \Bigl\vert \; T \Big\{ 
\phi(x) \; \phi(x') 
\Bigr\} \; \Bigr\vert 0 \Bigr\rangle =& \;
i\Delta_A^s(x;x') \; \frac{s}{1+s}  
+ i\Delta_C^s(x;x') \; \frac{1}{1+s} 
\;\; , &(3.38g) }$$
To assemble expresssions (3.38) in covariant form, we use (3.18)
to derive an identity:
$$\eqalignno{
\delta_{\mu}^{~(\alpha} \; \delta_{\nu}^{~\beta)} &=
{\overline \delta}_{\mu}^{~(\alpha} \; 
{\overline \delta}_{\nu}^{~\beta)} 
+ 2 \delta_{(\mu}^{~~0} \; {\overline \delta}_{\nu)}^{~(\alpha} \;
\delta_{0}^{~\beta)}
+ \delta_{\mu}^{~0} \; \delta_{\nu}^{~0} \;
\delta_{0}^{~\alpha} \; \delta_{0}^{~\beta}
\qquad \Longleftrightarrow \qquad &(3.39a) \cr
\psi_{\mu \nu} &=
\psi_{{\overline \mu}{\overline \nu}}
+ 2 \delta_{(\mu}^{~~0} \; \psi_{{\overline \nu \;})0}
+ \delta_{\mu}^{~0} \; \delta_{\nu}^{~0} \; \psi_{00}
\;\; , &(3.39b) }$$
which allows us to write the pseudo-graviton propagator as:
$$\eqalignno{
i\Bigl[ {_{\mu \nu}}\Delta_{\rho \sigma}^s \Bigr](x;x') =&
\; i\Delta_A^s(x;x') \; \Bigl\{ \;
\Bigl[ {_{\mu \nu}}T_{\rho \sigma}^A \Bigr] 
+ \frac{1}{1+s} 
\Bigl[ {_{\mu \nu}}T_{\rho \sigma}^C \Bigr]
\; \Bigr\}
+ i\Delta_B^s(x;x') \;
\Bigl[ {_{\mu \nu}}T_{\rho \sigma}^B \Bigr] \cr
&+ i\Delta_C^s(x;x') \;
\frac{s}{1+s}
\Bigl[ {_{\mu \nu}}T_{\rho \sigma}^C \Bigr] 
\;\; , &(3.40) }$$
where:
$$\eqalignno{
\Bigl[ {_{\mu \nu}}T_{\rho \sigma}^A \Bigr] &\equiv
2 \Bigl[ 
{\overline \eta}_{\mu(\rho} \; {\overline \eta}_{\sigma)\nu}
- {\overline \eta}_{\mu \nu} \; {\overline \eta}_{\rho \sigma}
\Bigr]
\;\; , &(3.41a) \cr
\Bigl[ {_{\mu \nu}}T_{\rho \sigma}^B \Bigr] &\equiv
-4 \delta_{(\mu}^{~~0} \; {\overline \eta}_{\nu)(\rho} \; 
\delta_{\sigma)}^{~~0}
\;\; , &(3.41b) \cr
\Bigl[ {_{\mu \nu}}T_{\rho \sigma}^C \Bigr] &\equiv
\Bigl[ 
{\overline \eta}_{\mu \nu} + 
\delta_{\mu}^{~0} \; \delta_{\nu}^{~0}
\Bigr] \; \Bigl[
{\overline \eta}_{\rho \sigma} + 
\delta_{\rho}^{~0} \; \delta_{\sigma}^{~0}
\Bigr]
\;\; . &(3.41c) }$$
There is also the mixed propagator which equals:
$$i\Bigl[ {_{\mu \nu}}\Delta^s \Bigr](x;x') = 
\Bigl( -i \Delta_A^s + i \Delta_C^s \Bigr) 
\; \frac{\sqrt s}{1+s} \;
\Bigl[ {\overline \eta}_{\mu \nu} + 
\delta_{\mu}^{~0} \; \delta_{\nu}^{~0} \Bigr] 
\;\; . \eqno(3.42)$$

\vfil\eject
{\it 3.4 The Explicit Form of the Propagators.}

The condition (3.32) on the vacuum and the free field expansions
(3.31) imply that:
$$\eqalignno{
i\Delta_I(x;x') = \int &
{d^3k \over (2\pi)^3} \; \exp(-\epsilon k) \;
\exp \Bigl[ i{\vec k} \cdot ({\vec x} - {\vec x'}) \Bigr] \;
\Bigl\{ \; \theta(\eta - \eta') \; 
\Psi(\eta, k, I) \; \Psi^{\star}(\eta', k, I) \cr
&+ \theta(\eta' - \eta) \; 
\Psi^{\star}(\eta, k, I) \; \Psi(\eta', k, I) \; \Bigr\}
\qquad , \qquad I=A,B,C
\;\; . &(3.43a) }$$ 
It is elementary to perform the angular integrations:
$$\eqalignno{
i\Delta_I(x;x') = 
{1 \over 2\pi^2 \Delta x} & \int_0^{+\infty} 
dk \; \exp(-\epsilon k) \; k \; \sin(k \Delta x) \;
\Bigl\{ \; \theta(\Delta\eta) \; 
\Psi(\eta, k, I) \; \Psi^{\star}(\eta', k, I) \cr
&+ \theta(- \Delta\eta) \; 
\Psi^{\star}(\eta, k, I) \; \Psi(\eta', k, I) \; \Bigr\}
\qquad , \qquad I=A,B,C
\;\; , \qquad\qquad &(3.43b) }$$ 
where $\Delta x \equiv \Vert {\vec x} - {\vec x'} \Vert$ and 
$\Delta \eta \equiv \eta - \eta'$. Notice that we have included 
the traditional ultraviolet convergence factor $\exp(-\epsilon k)$ 
which serves as an ultraviolet mode cutoff.

The actual computation of the propagators is a three-step process.
First, we solve equation (3.34) to obtain the mode functions
$\Psi(\eta, k, I)$. Then, we evaluate the propagators $i\Delta_I^s
(x;x')$ by inserting the respective mode functions in (3.43). In
the final step, we simply substitute the derived expressions for 
$i\Delta_I^s(x;x')$ in (3.38g) , (3.40) and (3.42) to arrive at the
scalar, pseudo-graviton and mixed propagators respectively.

{\it Step I.} Consider, to begin with, the ``$A$'' modes equation:
$${\rm D}_A^s \; f(\eta, {\vec x}) = 0
\qquad \Longrightarrow \qquad
\Omega \Bigl[ \;
-\partial_{\eta}^2 + \nabla^2 + 
\frac{2s^2 -s}{(1-s)^2} \; {1 \over \eta^2}
\; \Bigr] \Omega \; f(\eta, {\vec x}) = 0
\;\; . \eqno(3.44)$$ 
Now rewrite (3.44) as follows:
$$\Omega^2 \Bigl[ \;
-\partial_{\eta}^2 + \nabla^2  
-\frac{2s}{1-s} \; {1 \over \eta} \; \partial_{\eta}
\; \Bigr] \; f(\eta, {\vec x}) = 0
\;\; , \eqno(3.45)$$ 
and Fourier transform in momentum space:
$$\Bigl[ \; \partial_{\eta}^2 +  
\frac{2s}{1-s} \; {1 \over \eta} \; \partial_{\eta} + k^2 
\; \Bigr] \; {\widetilde f}(\eta, {\vec k}) = 0
\;\; , \eqno(3.46)$$ 
This can be cast in the form of the Bessel equation:
$$\Bigl( \; {d^2 \over dy^2} 
+ {1 \over y} \; {d \over dy} + 1 - {\nu^2 \over y^2}
\; \Bigr) \; Z_{\nu} = 0
\;\; , \eqno(3.47)$$
by substituting:
$$y \equiv k\eta 
\qquad ; \qquad 
g(y) \equiv y^w \; G(y) \equiv {\widetilde f}(\eta, {\vec k})
\;\; , \eqno(3.48)$$
so that (3.46) becomes:
$$y^w \Bigl[ \;
{d^2 \over dy^2} 
+ (\frac{2s}{1-s} + 2w) \; {1 \over y} \; {d \over dy} 
+ 1 + w \; (\frac{2s}{1-s} + w - 1) \; {1 \over y^2}
\; \Bigr] \; G(y) = 0
\; \; . \eqno(3.49)$$
This is the Bessel equation for $w=\frac{1-3s}{2(1-s)}$ . The two
solutions to (3.49) are Hankel functions:
$$G^{(1,2)}(y) = H_{\nu_A}^{(1,2)}(y)
\qquad , \qquad
\nu_A = \frac12 \Bigl\vert \frac{1-3s}{1-s} \Bigr\vert
\;\; , \eqno(3.50)$$
and are complex conjugates of each other: 
$$H_{\nu}^{(2)}(y) = H_{\nu}^{(1)\;\star}(y)
\;\; . \eqno(3.51)$$

A similar analysis is applied to the case of the ``$B$'' modes:
$${\rm D}_B^s \; f(\eta, {\vec x}) = 0
\qquad \Longrightarrow \qquad
\Omega \Bigl[ \;
-\partial_{\eta}^2 + \nabla^2 + 
\frac{s}{(1-s)^2} \; {1 \over \eta^2}
\; \Bigr] \Omega \; f(\eta, {\vec x}) = 0
\;\; . \eqno(3.52)$$ 
In momentum space equation (3.52) becomes:
$$\Bigl[ \; \partial_{\eta}^2 
+\frac{2s}{1-s} \; {1 \over \eta} \; \partial_{\eta} 
-\frac{2s}{1-s} \; {1 \over \eta^2} + k^2 
\; \Bigr] \; {\widetilde f}(\eta, {\vec k}) = 0
\;\; , \eqno(3.53)$$ 
which is again equal -- using (3.48) -- to a Bessel equation when
$w = \frac{1-3s}{2(1-s)}$ :
$$y^w \Bigl[ \;
{d^2 \over dy^2} 
+ (\frac{2s}{1-s} + 2w) \; {1 \over y} \; {d \over dy} + 1 + 
\Bigl( \; w \; (\frac{2s}{1-s} + w - 1) - \frac{2s}{1-s} \;
\Bigr) \; {1 \over y^2}
\; \Bigr] \; G(y) = 0
\; \; . \eqno(3.54)$$
The following set of complex conjugate solutions are implied:
$$G^{(1,2)}(y) = H_{\nu_B}^{(1,2)}(y)
\qquad , \qquad
\nu_B = \frac12 \Bigl\vert \frac{1+s}{1-s} \Bigr\vert
\;\; . \eqno(3.55)$$

Finally, the ``$C$'' modes lead to the following equation:
$$\Bigl[ \; \partial_{\eta}^2 +  
\frac{2s}{1-s} \; {1 \over \eta} \; \partial_{\eta} 
-\frac{2(1+s)}{1-s} \; {1 \over \eta^2} + k^2 
\; \Bigr] \; {\widetilde f}(\eta, {\vec k}) = 0
\;\; . \eqno(3.56)$$ 
The resulting Bessel equation is:
$$y^w \Bigl[ \;
{d^2 \over dy^2} 
+ (\frac{2s}{1-s} + 2w) \; {1 \over y} \; {d \over dy} + 1 + 
\Bigl( \; w \; (\frac{2s}{1-s} + w - 1) - \frac{2(1+s)}{1-s} \;
\Bigr) \; {1 \over y^2}
\; \Bigr] \; G(y) = 0
\; \; , \eqno(3.57)$$
where $w = \frac{1-3s}{2(1-s)}$. The solutions are:
$$G^{(1,2)}(y) = H_{\nu_C}^{(1,2)}(y)
\qquad , \qquad
\nu_C = \frac12 \Bigl\vert \frac{3-s}{1-s} \Bigr\vert
\;\; . \eqno(3.58)$$

To obtain the expressions for the mode functions, we must identify
the correspondence between $\Psi$ and $\Psi^{\star}$ -- as defined
in (3.31) -- and the solutions $H_{\nu}^{(1)}$ and $H_{\nu}^{(2)}$.
This can be accomplished by requiring the normalizability of the 
vacuum. A useful asymptotic expansion of the two solutions is:
$$\eqalignno{
H_{\nu}^{(1)}(y) &=
{\sqrt {2 \over \pi y}} \; \exp \Bigl[ \;
i(y - \frac{\pi}{2} \nu - \frac{\pi}{2}) \; \Bigr] \;
\Bigl( \; 1 + O(y^{-1}) \; \Bigr)
\qquad , \qquad y \gg 1
\;\; , &(3.59a) \cr
H_{\nu}^{(2)}(y) &=
{\sqrt {2 \over \pi y}} \; \exp \Bigl[ \;
-i(y - \frac{\pi}{2} \nu - \frac{\pi}{2}) \; \Bigr] \;
\Bigl( \; 1 + O(y^{-1}) \; \Bigr)
\qquad , \qquad y \gg 1
\;\; , \qquad &(3.59b) }$$
Because of the exponential behaviour seen in (3.59) , we must 
associate $\Psi(\eta, k, I)$ -- the coefficient function of the 
annihilation operator in (3.32) -- with $H_{\nu_I}^{(2)}(k\eta)$ 
and $\Psi^{\star}(\eta, k, I)$ with $H_{\nu_I}^{(1)}(k\eta)$. 
After we account for the normalization $N$ and the transformation 
(3.48) , we get:
$$\eqalignno{
\Psi(\eta, k, I) &=
N(k) \; (k\eta)^w \; H_{\nu_I}^{(2)}(k\eta)
\qquad , \qquad I=A,B,C
\;\; , &(3.60a) \cr
\Psi^{\star}(\eta, k, I) &=
N(k) \; (k\eta)^w \; H_{\nu_I}^{(1)}(k\eta)
\qquad , \qquad I=A,B,C
\;\; . &(3.60b) }$$
The alternate connection, namely $\Psi \leftrightarrow H^{(1)}$
and $\Psi^{\star} \leftrightarrow H^{(2)}$, would lead to a vacuum
state that is not normalizable.
\footnote{*}{\tenpoint The normalizable vacuum state that follows
from prescription (3.60) is the Bunch-Davies vacuum [4]. Although 
this works fine for power laws, there is an ambiguity for a generic 
scale factor. Since the solutions in the latter case are not known
it is not clear whether a proper vacuum state exists in general.}

The normalization of the mode functions comes from (3.33). When we
take into account relations (3.60) and the Hankel function identity:
$$H_{\nu}^{(2)}(y) \;\; {d \over dy} H_{\nu}^{(1)}(y)
- H_{\nu}^{(1)}(y) \;\; {d \over dy} H_{\nu}^{(2)}(y)
= {4i \over \pi y}
\;\; , \eqno(3.61)$$
equation (3.33) gives:
$$N(k) = \frac12 {\sqrt \pi} \;\;
\eta_0^{\frac{s}{1-s}} \;\; k^{-\frac{1-3s}{2(1-s)}}
\;\; , \eqno(3.62)$$
so that: 
$$\eqalignno{
\Psi(\eta, k, I) &=
\frac12 {\sqrt {\pi\eta}} \;\; \Omega^{-1}(\eta) \;\; 
H_{\nu_I}^{(2)}(k\eta)
\qquad , \qquad I=A,B,C
\;\; , &(3.63a) \cr
\Psi^{\star}(\eta, k, I) &=
\frac12 {\sqrt {\pi\eta}} \;\; \Omega^{-1}(\eta) \;\; 
H_{\nu_I}^{(1)}(k\eta)
\qquad , \qquad I=A,B,C
\;\; . &(3.63b) }$$
The basic parameter values of the most physically interesting
power laws are displayed in Table 1. The kinds of Hankel functions
generated by inflation, matter, radiation and flatness are presented
in Table 2 and their functional form in Table 3. The explicit
expressions which the mode functions take in these cases are shown
in Table 4. In the case of the inflating universe ($s = +\infty$)
they are in agreement with previously obtained results [5].

\vskip 1.5cm 

\vbox{\tabskip=0pt \offinterlineskip
\def\tablerule{\noalign{\hrule}}
\halign to480pt {\strut#& \vrule#\tabskip=1em plus2em& 
\hfil#& \vrule#& \hfil#\hfil& \vrule#& \hfil#& \vrule#& \hfil#\hfil& 
\vrule#& \hfil#\hfil&
\vrule#\tabskip=0pt\cr
\tablerule
\omit&height4pt&\omit&&\omit&&\omit&&\omit&&\omit&\cr
&&\omit\hidewidth {}\hidewidth&&
\omit\hidewidth {Flat}\hidewidth&& 
\omit\hidewidth {Radiation}\hidewidth&& 
\omit\hidewidth {Matter}\hidewidth&& 
\omit\hidewidth {Inflation} 
\hidewidth&\cr
\omit&height4pt&\omit&&\omit&&\omit&&\omit&&\omit&\cr
\tablerule
\omit&height3pt&\omit&&\omit&&\omit&&\omit&&\omit&\cr
&& $s$ && $0$ && $\frac12$ && $\frac23$ && $+\infty$ &\cr
\omit&height3pt&\omit&&\omit&&\omit&&\omit&&\omit&\cr
\tablerule
\omit&height3pt&\omit&&\omit&&\omit&&\omit&&\omit&\cr
&& $\Omega(\eta)$ && $1$ && $\frac{\eta}{\eta_0}$ 
&& $\frac{\eta^2}{\eta_0^2}$ && $-\frac{\eta_0}{\eta}$ &\cr
\omit&height3pt&\omit&&\omit&&\omit&&\omit&&\omit&\cr
\tablerule}}

\vskip 0.2cm

{\bf Table~1:} {\ninepoint The power law parameter $s$ and scale factor 
  $\Omega(\eta)$ for some special spacetimes.}

\vskip 1.5cm 

\vbox{\tabskip=0pt \offinterlineskip
\def\tablerule{\noalign{\hrule}}
\halign to480pt {\strut#& \vrule#\tabskip=1em plus2em& 
\hfil#& \vrule#& \hfil#\hfil& \vrule#& \hfil#& \vrule#& \hfil#\hfil& 
\vrule#& \hfil#\hfil&
\vrule#\tabskip=0pt\cr
\tablerule
\omit&height4pt&\omit&&\omit&&\omit&&\omit&&\omit&\cr
&&\omit\hidewidth {$H_{\nu_I}^{(1,2)}(k \eta)\;$ Index}\hidewidth&&
\omit\hidewidth {$s = 0$}\hidewidth&& 
\omit\hidewidth {$s = \frac12$}\hidewidth&& 
\omit\hidewidth {$s = \frac23$}\hidewidth&& 
\omit\hidewidth {$s = +\infty$} 
\hidewidth&\cr
\omit&height4pt&\omit&&\omit&&\omit&&\omit&&\omit&\cr
\tablerule
\omit&height3pt&\omit&&\omit&&\omit&&\omit&&\omit&\cr
&& $\nu_A$ && $\frac12$ && $\frac12$ && $\frac32$ && $\frac32$ &\cr
\omit&height3pt&\omit&&\omit&&\omit&&\omit&&\omit&\cr
\tablerule
\omit&height3pt&\omit&&\omit&&\omit&&\omit&&\omit&\cr
&& $\nu_B$ && $\frac12$ && $\frac32$ && $\frac52$ && $\frac12$ &\cr
\omit&height3pt&\omit&&\omit&&\omit&&\omit&&\omit&\cr
\tablerule
\omit&height3pt&\omit&&\omit&&\omit&&\omit&&\omit&\cr
&& $\nu_C$ && $\frac32$ && $\frac52$ && $\frac72$ && $\frac12$ &\cr
\omit&height3pt&\omit&&\omit&&\omit&&\omit&&\omit&\cr
\tablerule}}

\vskip 0.2cm

{\bf Table~2:} {\ninepoint The value of the index of the Hankel functions
  for the special power law scale factors.}

\vfil\eject

\vbox{\tabskip=0pt \offinterlineskip
\def\tablerule{\noalign{\hrule}}
\halign to480pt {\strut#& \vrule#\tabskip=1em plus2em& 
\hfil#& \vrule#& \hfil#\hfil& 
\vrule#\tabskip=0pt\cr
\tablerule
\omit&height4pt&\omit&&\omit&\cr
&&\omit\hidewidth {$H_{\nu}^{(2)}(k \eta)\;$ Index}\hidewidth&&
\omit\hidewidth {Functional Form of $\;H_{\nu}^{(2)}(k \eta)$}
\hidewidth&\cr
\omit&height4pt&\omit&&\omit&\cr
\tablerule
\omit&height3pt&\omit&&\omit&\cr
&& $\nu = \frac12$ && 
$i \sqrt{2 \over {\pi k \eta}} \; \exp{[ -i k \eta]}$ &\cr
\omit&height3pt&\omit&&\omit&\cr
\tablerule
\omit&height3pt&\omit&&\omit&\cr
&& $\nu = \frac32$ && 
$- \sqrt{2 \over {\pi k \eta}} \; \exp{[ -i k \eta]} \;
\Bigl( \; 1 - {i \over {k \eta}} \; \Bigr)$ &\cr
\omit&height3pt&\omit&&\omit&\cr
\tablerule
\omit&height3pt&\omit&&\omit&\cr
&& $\nu = \frac52$ && 
$-i \sqrt{2 \over {\pi k \eta}} \; \exp{[ -i k \eta]} \;
\Bigl( \; 1 - {3i \over {k \eta}} - {3 \over (k \eta)^2} \; \Bigr)$ &\cr
\omit&height3pt&\omit&&\omit&\cr
\tablerule
\omit&height3pt&\omit&&\omit&\cr
&& $\nu = \frac72$ && 
$\sqrt{2 \over {\pi k \eta}} \; \exp{[ -i k \eta]} \;
\Bigl( \; 1 - {6i \over {k \eta}} - {15 \over (k \eta)^2} 
+ {15i \over (k \eta)^3} \; \Bigr)$ &\cr
\omit&height3pt&\omit&&\omit&\cr
\tablerule}}

\vskip 0.2cm

{\bf Table~3:} {\ninepoint The functional form of $H_{\nu}^{(2)}$ for
  the needed half-integer values of the index $\nu$; the form of
\vskip -16pt \noindent \hglue 2.85truecm $H_{\nu}^{(1)}$ is obtained 
by complex conjugation.} 

\vskip 1.5cm

\vbox{\tabskip=0pt \offinterlineskip
\def\tablerule{\noalign{\hrule}}
\halign to480pt {\strut#& \vrule#\tabskip=1em plus2em& 
\hfil#& \vrule#& \hfil#\hfil& \vrule#& \hfil#\hfil& 
\vrule#\tabskip=0pt\cr
\tablerule
\omit&height4pt&\omit&&\omit&&\omit&\cr
&&\omit\hidewidth {Power Law}\hidewidth&&
\omit\hidewidth {Functional Form} \hidewidth&& 
\omit\hidewidth {Mode Function} 
\hidewidth&\cr
\omit&height4pt&\omit&&\omit&&\omit&\cr
\tablerule
\omit&height3pt&\omit&&\omit&&\omit&\cr
&& $s = 0$ && ${i \over \sqrt {2k}} \; \exp{[-ik \eta]}$ 
&& $\Psi(\eta, k, A) = \Psi(\eta, k, B)$ &\cr
\omit&height3pt&\omit&&\omit&&\omit&\cr
\tablerule
\omit&height3pt&\omit&&\omit&&\omit&\cr
&& $s = 0$ && ${i \over \sqrt {2k^3}} \; {1 \over \eta} \;
( \; 1 + i k \eta \; ) \; \exp{[-ik \eta]}$ 
&& $\Psi(\eta, k, C)$ &\cr
\omit&height3pt&\omit&&\omit&&\omit&\cr
\tablerule
\omit&height3pt&\omit&&\omit&&\omit&\cr
&& $s = \frac12$ && ${i \over \sqrt {2k}} \; {\eta_0 \over \eta} 
\; \exp{[-ik \eta]}$ 
&& $\Psi(\eta, k, A)$ &\cr
\omit&height3pt&\omit&&\omit&&\omit&\cr
\tablerule
\omit&height3pt&\omit&&\omit&&\omit&\cr
&& $s = \frac12$ && ${i \over \sqrt {2k^3}} \; {\eta_0 \over \eta^2} \;
( \; 1 + i k \eta \; ) \; \exp{[-ik \eta]}$ 
&& $\Psi(\eta, k, B)$ &\cr
\omit&height3pt&\omit&&\omit&&\omit&\cr
\tablerule
\omit&height3pt&\omit&&\omit&&\omit&\cr
&& $s = \frac12$ && ${i \over \sqrt{2k}} \; {\eta_0 \over \eta} \;
\Bigl( \; {3 \over (k \eta)^2} + {3i \over k \eta} - 1 
\; \Bigr) \; \exp{[-ik \eta]}$ 
&& $\Psi(\eta, k, C)$ &\cr
\omit&height3pt&\omit&&\omit&&\omit&\cr
\tablerule
\omit&height3pt&\omit&&\omit&&\omit&\cr
&& $s = \frac23$ && ${i \over \sqrt{2k^3}} \; {\eta_0^2 \over \eta^3} \; 
( \; 1 + i k \eta \; ) \; \exp{[-ik \eta]}$ 
&& $\Psi(\eta, k, A)$ &\cr
\omit&height3pt&\omit&&\omit&&\omit&\cr
\tablerule
\omit&height3pt&\omit&&\omit&&\omit&\cr
&& $s = \frac23$ && ${i \over \sqrt{2k}} \; {\eta_0^2 \over \eta^2} \; 
\Bigl( \; {3 \over (k \eta)^2} + {3i \over k \eta} - 1 
\; \Bigr) \; \exp{[-ik \eta]}$ 
&& $\Psi(\eta, k, B)$ &\cr
\omit&height3pt&\omit&&\omit&&\omit&\cr
\tablerule
\omit&height3pt&\omit&&\omit&&\omit&\cr
&& $s = \frac23$ && ${1 \over \sqrt{2k}} \; {\eta_0^2 \over \eta^2} \; 
\Bigl( \; 1 - {6i \over k \eta} - {15 \over (k \eta)^2} + 
{15i \over (k \eta)^3}
\; \Bigr) \; \exp{[-ik \eta]}$ 
&& $\Psi(\eta, k, C)$ &\cr
\omit&height3pt&\omit&&\omit&&\omit&\cr
\tablerule
\omit&height3pt&\omit&&\omit&&\omit&\cr
&& $s = +\infty$ && $-{i \over \sqrt {2k^3}} \; {1 \over \eta_0} \;
( \; 1 + i k \eta \; ) \; \exp{[-ik \eta]}$ 
&& $\Psi(\eta, k, A)$ &\cr
\omit&height3pt&\omit&&\omit&&\omit&\cr
\tablerule
\omit&height3pt&\omit&&\omit&&\omit&\cr
&& $s = +\infty$ && $-{i \over \sqrt {2k}} \; {\eta \over \eta_0} \; 
\exp{[-ik \eta]}$ 
&& $\Psi(\eta, k, B) = \Psi(\eta, k, C)$ &\cr
\omit&height3pt&\omit&&\omit&&\omit&\cr
\tablerule}}

\vskip 0.2cm

{\bf Table~4:} {\ninepoint The form of $\Psi(\eta, k, I)$ for the 
  special power law scale factors; the form of $\Psi^*(\eta, k, I)$ 
  is
\vskip -16pt \noindent \hglue 2.85truecm obtained by complex conjugation.}

\vfil\eject

{\it Step II.} We proceed to calculate the propagators $i\Delta_I^s 
(x;x')$ by substituting relations (3.63) in (3.43b) to obtain:
$$\eqalignno{
i\Delta_{\nu_I}(x;x') = &
{\sqrt{\eta \eta'} \; \Omega^{-1}(\eta) \;\; \Omega^{-1}(\eta')
\over 8\pi \Delta x} 
\int_0^{+\infty} 
dk \; \exp(-\epsilon k) \; k \; \sin(k \Delta x) \cr
& \Bigl\{ \; \theta(\Delta\eta) \;\; 
H_{\nu_I}^{(2)}(k\eta) \;\; H_{\nu_I}^{(1)}(k\eta') 
+ \theta(- \Delta\eta) \;\; 
H_{\nu_I}^{(1)}(k\eta) \;\; H_{\nu_I}^{(2)}(k\eta') 
\; \Bigr\}
\;\; . \qquad &(3.64) }$$ 
An infrared divergence coming from the lower limit of integration
may exist here. When present, it is due to the infinite size
universe imposed by the integral approximation to the mode 
sums we have assumed. This divergence is not physical and has 
appeared in previous studies of the graviton propagator in de 
Sitter spacetime [6]. 
\footnote{*}{\tenpoint For de Sitter spacetime, this does not 
preclude the appearance of infrared divergences in physical
quantities [7,8].} On our finite spatial manifold the propagator
is a mode sum over discrete momenta, so it is perfectly well
defined. The integral approximation to this mode sum has lower
limit $k_0 \sim {\rm R}^{-1}$ [8].

It is interesting to examine the range of power laws for which 
this infrared divergence exists in the basic integral (3.64). 
The dominant behaviour of the Hankel functions for small $k$:
$$H_{\nu}^{(2)}(y) \qquad \longmapsto \qquad
{(\nu -1)! \; 2^{\nu} \over i\pi} \; y^{-\nu}
\qquad , \qquad k \ll 1
\;\; , \eqno(3.65)$$
translates to the following behaviour for the propagators:
$$i\Delta_{\nu_I}(x;x') \qquad \approx \qquad
\int_0^{+\infty} dk \; k^{2(1 - \nu_I)}
\qquad , \qquad k \ll 1
\;\; , \eqno(3.66)$$
and, therefore, the following condition for the presence of an
infrared divergence:
$$1 - \nu_I \; \leq \; -\frac12 
\;\; . \eqno(3.67)$$
For the ``A'' modes (3.67) implies:
$$\nu_A = \frac12 \Bigl\vert \frac{1-3s}{1-s} \Bigr\vert
\qquad \Longrightarrow \qquad s \geq \frac23
\;\; , \eqno(3.68)$$
while for the ``B'' and ``C'' modes we get:
$$\eqalignno{
\nu_B = \frac12 \Bigl\vert \frac{1+s}{1-s} \Bigr\vert
\qquad &\Longrightarrow \qquad 
\frac12 \leq s \leq 2
\;\; , &(3.69a) \cr
\nu_C = \frac12 \Bigl\vert \frac{3-s}{1-s} \Bigr\vert
\qquad &\Longrightarrow \qquad 0 \leq s \leq \frac32 
\;\; . &(3.69b) }$$ 
Since the physical polarizations of the graviton are ``A'' modes,
the corresponding crossover value $s_{\rm crit} = \frac23$ may have
some indirect physical significance [9].

We now turn to the computation of the propagators $i\Delta_{\nu_I}
(x;x')$ for the power laws of interest. As Table 2 indicates, four
Hankel functions are needed. Substituting the expressions of Table 3
in (3.64), we get for $\nu = \frac12$ :
$$\eqalignno{
i\Delta_{\frac12}(x;x') &= 
{1 \over 4\pi^2 \; \Omega(\eta) \; \Omega(\eta') \; \Delta x} 
\; \int_0^{+\infty} dk \; \sin(k \Delta x) \;
\exp \Bigl[ -ik (\vert\Delta\eta\vert - i\epsilon) \Bigr] \cr
&= 
{1 \over 4\pi^2 \; \Omega(\eta) \; \Omega(\eta')} \;\;
{1 \over (\Delta x - \vert\Delta\eta\vert + i\epsilon) \;
(\Delta x + \vert\Delta\eta\vert - i\epsilon)}
\;\; , &(3.70) }$$ 
for $\nu = \frac32$ :
$$\eqalignno{
i\Delta_{\frac32}(x;x') &= 
{1 \over 4\pi^2 \; \Omega(\eta) \; \Omega(\eta') \; \Delta x} 
\; \int_{k_0}^{+\infty} dk \; \sin(k \Delta x) \;
\exp \Bigl[ -ik (\vert\Delta\eta\vert - i\epsilon) \Bigr] \;
\Biggl[ \; 1 + {1 + ik\vert\Delta\eta\vert 
\over k^2 \; \eta \; \eta'} \; \Biggr] \cr
&= \;
i\Delta_{\frac12}(x;x') -
{1 \over 8\pi^2 \; \Omega(\eta) \; \eta \;\; 
\Omega(\eta') \; \eta'} \;
\Bigl\{ \; -2 +
{\rm Ei}\Bigl[ ik_0 (\Delta x - \vert\Delta\eta\vert + i\epsilon) 
\Bigr] \cr
&\qquad\qquad\qquad\qquad\qquad\qquad\qquad\qquad\quad\;
+{\rm Ei}\Bigl[ -ik_0 (\Delta x + \vert\Delta\eta\vert - i\epsilon) 
\Bigr] \; \Bigr\}
\;\; , \qquad &(3.71) }$$ 
for $\nu = \frac52$ :
$$\eqalignno{
i\Delta_{\frac52}(x;x') &= 
{1 \over 4\pi^2 \; \Omega(\eta) \; \Omega(\eta') \; \Delta x} 
\; \int_{k_0}^{+\infty} dk \; \sin(k \Delta x) \;
\exp \Bigl[ -ik (\vert\Delta\eta\vert - i\epsilon) \Bigr] \cr
&\qquad\qquad\qquad\qquad\qquad\qquad\quad
\Biggl[ \; 1 + 
3{1 + ik\vert\Delta\eta\vert \over k^2 \; \eta \; \eta'} +
{9 + 9ik\vert\Delta\eta\vert - 3k^2 (\Delta\eta)^2 
\over k^4 \; \eta^2 \; {\eta'}^2}
\; \Biggr] \cr
&= \;
-2i\Delta_{\frac12}(x;x') + 3i\Delta_{\frac32}(x;x') \cr
&\quad\; +
{1 \over 8\pi^2 \; \Omega(\eta) \; \eta^2 \;\; 
\Omega(\eta') \; {\eta'}^2} \;
\Biggl\{ \;
9k_0^{-2} - \frac{11}2 (\Delta x)^2 + \frac92 (\Delta\eta)^2
+ \frac32 \Bigl[ (\Delta x)^2 - (\Delta\eta)^2 \Bigr] \cr
&\qquad\quad \times 
\Bigl\{ \; 
{\rm Ei}\Bigl[ ik_0 (\Delta x - \vert\Delta\eta\vert + i\epsilon) 
\Bigr] + 
{\rm Ei}\Bigl[ -ik_0 (\Delta x + \vert\Delta\eta\vert - i\epsilon) 
\Bigr] \; \Bigr\} \; \Biggr\}
\;\; , \qquad &(3.72) }$$ 
and for $\nu = \frac72$ :
$$\eqalignno{
i\Delta_{\frac72}(x;x') &= 
{1 \over 4\pi^2 \; \Omega(\eta) \; \Omega(\eta') \; \Delta x} 
\; \int_{k_0}^{+\infty} dk \; \sin(k \Delta x) \;
\exp \Bigl[ -ik (\vert\Delta\eta\vert - i\epsilon) \Bigr] \cr
&\qquad\qquad\qquad\qquad\qquad\qquad\quad
\Biggl[ \; 1 + 
6{1 + ik\vert\Delta\eta\vert \over k^2 \; \eta \; \eta'} +
{45 + 45ik\vert\Delta\eta\vert - 15k^2 (\Delta\eta)^2 
\over k^4 \; \eta^2 \; {\eta'}^2} \cr
&\qquad\qquad\qquad\qquad\qquad\qquad\quad\quad\;\;
+ {225 + 225ik\vert\Delta\eta\vert - 90k^2 \Delta\eta^2 -
15ik^3 \vert\Delta\eta\vert^3 \over
k^6 \; \eta^3 \; {\eta'}^3}
\; \Biggr] \cr 
&= \;
5i\Delta_{\frac12}(x;x') - 9i\Delta_{\frac32}(x;x') 
+ 5i\Delta_{\frac52}(x;x') \cr
&\quad + 
{1 \over 16\pi^2 \; \Omega(\eta) \; \eta^3 \;\; 
\Omega(\eta') \; {\eta'}^3} \;
\Biggl\{ \; 225 k_0^{-4} 
- \Bigl[ 75(\Delta x)^2 - 45(\Delta\eta)^2 \Bigr]
k_0^{-2} \cr
&\quad\quad + 
\Bigl[ \frac{137}8 (\Delta x)^4 
- \frac{125}4 (\Delta x)^2 (\Delta\eta)^2
+ \frac{105}8 (\Delta\eta)^4 \Bigr]
- \frac{15}4 \Bigl( (\Delta x)^2 - (\Delta\eta)^2 \Bigr)^2 \cr
&\quad\quad\quad \times 
\Bigl\{ \; 
{\rm Ei}\Bigl[ ik_0 (\Delta x - \vert\Delta\eta\vert + i\epsilon) 
\Bigr] + 
{\rm Ei}\Bigl[ -ik_0 (\Delta x + \vert\Delta\eta\vert - i\epsilon) 
\Bigr] \; \Bigr\} \; \Biggr\}
\;\; , \qquad &(3.73) }$$ 

The integration technique consists of performing as many integrations 
by parts as needed to reduce the original expression to the following 
basic integrals:
$$\eqalignno{
\int_0^{+\infty} dk \; {\rm sin}(k \Delta x) \;
\exp{ \Bigl[ -ik(\vert\Delta\eta\vert - i\epsilon) \Bigr] } &=
{\Delta x \over (\Delta x - \vert\Delta\eta\vert + i\epsilon) \;
(\Delta x + \vert\Delta\eta\vert - i\epsilon)}
\;\; , &(3.74) \cr
\int_{k_0}^{+\infty} {dk \over k} \; {\rm cos}(k \Delta x) \;
\exp{ \Bigl[ -ik(\vert\Delta\eta\vert - i\epsilon) \Bigr] } &=
-\frac12 \Bigl\{ \; {\rm Ei} \Bigl[ 
ik_0(\Delta x - \vert\Delta\eta\vert + i\epsilon) \Bigr] \cr
&\qquad\;\;\;\; + {\rm Ei} \Bigl[
-ik_0(\Delta x + \vert\Delta\eta\vert - i\epsilon) \Bigr] \;
\Bigr\} \;\; . \qquad\qquad &(3.75) }$$
The first of these does not exhibit an infrared divergence but the
second does and, therefore, its lower limit of integration is regulated 
to the value $k_0 \sim {\rm R}^{-1}$. The exponential integral function 
${\rm Ei}(x)$ is defined as [10]:
$${\rm Ei}(-x) = -\int_x^{+\infty} dt \; t^{-1} \; \exp{(-t)}
\;\; , \eqno(3.76)$$
and for small $x$ is very well approximated by:
$${\rm Ei}(-x) \approx \gamma + \ln x \qquad , \qquad x\ll 1
\;\; , \eqno(3.77)$$
where $\gamma$ stands for Euler's constant. It is legitimate to employ
this approximation in our case:
$$\eqalignno{
\int_{k_0}^{+\infty} {dk \over k} \; {\rm cos}(k \Delta x) \;
&\exp{ \Bigl[ -ik(\vert\Delta\eta\vert - i\epsilon) \Bigr] } \approx \cr 
&-\frac12  \Bigl\{ \; \ln \Bigl[  k_0^2
(\Delta x - \vert\Delta\eta\vert + i\epsilon) \; 
(\Delta x + \vert\Delta\eta\vert - i\epsilon) \Bigr] 
+ 2\gamma \; \Bigr\} 
\;\; , \qquad &(3.78) }$$
since we have assumed throughout that $k_0 \Delta x \ll 1$ and 
$k_0 \vert\Delta\eta\vert \ll 1$.

Using the approximation (3.78) and the definition:
$$(x - x')^2 \equiv 
(\Delta x - \vert\Delta\eta\vert + i\epsilon) \; 
(\Delta x + \vert\Delta\eta\vert - i\epsilon)  
\;\; , \eqno(3.79)$$
we can express $i\Delta_{\nu_I}(x;x')$ in a more economical form:
$$i\Delta_{\frac12}(x;x') = 
{1 \over 4\pi^2 \; \Omega(\eta) \; \Omega(\eta')} \;\;
{1 \over (x - x')^2}
\;\; , \eqno(3.80)$$ 
$$i\Delta_{\frac32}(x;x') = 
{1 \over 8\pi^2 \; \Omega(\eta) \; \eta \;\; 
\Omega(\eta') \; \eta'} \;
\Biggl\{ \; 
{2\eta \; \eta' \over (x -x')^2} 
- \ln \Bigl[ k_0^2 (x - x')^2 \Bigr]
- 2(\gamma - 1)
\; \Biggr\}
\;\; , \eqno(3.81)$$ 
$$\eqalignno{
i\Delta_{\frac52}(x;x') &= 
{1 \over 8\pi^2 \; \Omega(\eta) \; \eta^2 \;\; 
\Omega(\eta') \; {\eta'}^2} \;
\Biggl\{ \; 
{2\eta^2 \; {\eta'}^2 \over (x -x')^2} 
+ \frac32 \Bigl[ (x-x')^2 - 2\eta \; \eta' \Bigr] \;
\ln \Bigl[ k_0^2 (x - x')^2 \Bigr] \cr
&\qquad +
9k_0^{-2} - \frac{11}2 (\Delta x)^2 
+ \frac92 (\Delta\eta)^2 + 6\eta \; \eta'
+ 3\gamma \Bigl[ (x-x')^2 - 2\eta \; \eta' \Bigr]
\; \Biggr\}
\;\; , \qquad\qquad &(3.82) }$$ 
$$\eqalignno{
i\Delta_{\frac72}(x;x') &= 
{1 \over 16\pi^2 \; \Omega(\eta) \; \eta^3 \;\; 
\Omega(\eta') \; {\eta'}^3} \;
\Biggl\{ \; 
{4\eta^3 \; {\eta'}^3 \over (x -x')^2} 
- \Bigl[ \frac{15}4 (x-x')^4 - 5 (x - x')^2 \eta \; \eta' 
+ 12 \eta^2 \; {\eta'}^2 \Bigr] \cr
&\qquad\quad \times
\Bigl( \; \ln \Bigl[ k_0^2 (x - x')^2 \Bigr] 
+ 2\gamma \; \Bigr)
+ 225 k_0^{-4} 
- \Bigl[ 75(\Delta x)^2 - 45(\Delta\eta)^2 
- 90\eta \; \eta' \Bigr] k_0^{-2} \cr
&\qquad\qquad\qquad +
\Bigl[ \frac{137}8 (\Delta x)^4 
- \frac{125}4 (\Delta x)^2 (\Delta\eta)^2
+ \frac{105}8 (\Delta\eta)^4 \Bigr] \cr
&\qquad\qquad\qquad\qquad -
\Bigl[ 55(\Delta x)^2 - 45(\Delta\eta)^2
\Bigr] \eta \; \eta' 
+ 24\eta^2 \; {\eta'}^2 
\; \Biggr\}
\;\; . \qquad &(3.83) }$$ 

{\it Step III.} To obtain the explicit form of the scalar, 
pseudo-graviton and mixed propagators for the power laws of 
Table 1, we only need to appropriately substitute the above 
expressions in equations (3.38g) , (3.40) and (3.42) respectively. 
As an example we shall consider the case of de Sitter spacetime 
so that we can connect with previous work [3,8]. It corresponds
to $s \rightarrow +\infty$ and in this limit the scalar field
action becomes the cosmological constant term: 
$${\dot \varphi}^2 \rightarrow 0 
\qquad ; \qquad 
V(\varphi) \rightarrow (8\pi G) \; {\Lambda}^{-1}
\;\; , \eqno(3.84)$$
where $\Lambda = 3 \eta_0^{-2}$.
Consequently, only the pseudo-graviton propagator exists and it 
equals:
$$\eqalignno{
i\Bigl[ {_{\mu \nu}}\Delta_{\rho \sigma} \Bigr]_{s=+\infty}(x;x')
&= i\Delta_{\frac32}(x;x') \;
\Bigl[ {_{\mu \nu}}T^A_{\rho \sigma} \Bigr] + 
i\Delta_{\frac12}(x;x') \; 
\Biggl\{ \;
\Bigl[ {_{\mu \nu}}T^B_{\rho \sigma} \Bigr] +
\Bigl[ {_{\mu \nu}}T^C_{\rho \sigma} \Bigr] \;
\Biggr\} \cr
&= i\Delta_{\frac12}(x;x') \; 
\Biggl\{ \;
\Bigl[ {_{\mu \nu}}T^A_{\rho \sigma} \Bigr] + 
\Bigl[ {_{\mu \nu}}T^B_{\rho \sigma} \Bigr] +
\Bigl[ {_{\mu \nu}}T^C_{\rho \sigma} \Bigr] \;
\Biggr\} \cr
&\;\;\;\; - {1 \over 8\pi^2 \eta_0^2} \;
\Bigl\{ \; \ln \Bigl[ k_0^2 (x - x')^2 \Bigr]
+ 2(\gamma - 1) \; \Bigr\} \; 
\Bigl[ {_{\mu \nu}}T^A_{\rho \sigma} \Bigr] \cr
&= {1 \over 8\pi^2 \eta_0^2} \;
\Biggl\{ \;
{2\eta \; \eta' \over (x - x')^2} \;
\Biggl( \;
\Bigl[ {_{\mu \nu}}T^A_{\rho \sigma} \Bigr] + 
\Bigl[ {_{\mu \nu}}T^B_{\rho \sigma} \Bigr] +
\Bigl[ {_{\mu \nu}}T^C_{\rho \sigma} \Bigr] \;
\Biggr) \cr
&\qquad\qquad\;\;\;\;
- \Bigl\{ \; \ln \Bigl[ k_0^2 (x - x')^2 \Bigr]
+ 2(\gamma - 1) \; \Bigr\} \;
\Bigl[ {_{\mu \nu}}T^A_{\rho \sigma} \Bigr]
\; \Biggr\} 
\;\; , \qquad\qquad &(3.85) }$$ 
in agreement with previous results. By using the zero modes present
we can absorb the constant term $2(\gamma -1)$ and set $k_0 = 
\eta_0^{-1}$ [8]. However, this property is particular to the de 
Sitter spacetime and is not shared by the other power laws of 
interest. Finally, the combination of the tensor factors (3.41) that 
appears in (3.85) equals:
$$\Bigl[ {_{\mu \nu}}T^A_{\rho \sigma} \Bigr] + 
\Bigl[ {_{\mu \nu}}T^B_{\rho \sigma} \Bigr] +
\Bigl[ {_{\mu \nu}}T^C_{\rho \sigma} \Bigr] =
2\eta_{\mu ( \rho} \; \eta_{\sigma ) \nu} -
\eta_{\mu \nu} \; \eta_{\rho \sigma} 
\;\; . \eqno(3.86)$$

{\it 3.5 The Ghost Sector.}

The ghost Lagrangian is obtained by varying the gauge functional 
$F_{\mu}$ given by (3.14b). Under an infinitesimal coordinate change:
$${y'}^{\mu} = y^{\mu} + \kappa \; \omega^{\mu}(y) 
\;\; , \eqno(3.87)$$
the full metric transforms to:
$${g'}_{\mu \nu}(x) = 
{\partial y^{\rho} \over \partial y^{'~\mu}}(x) \;
{\partial y^{\sigma} \over \partial y^{'~\nu}}(x) \;
g_{\rho \sigma}\Bigl( {y'}^{-1}(x)\Bigr) 
\;\; , \eqno(3.88)$$
so that the infinitesimal variation of the pseudo-graviton field is:
$$\eqalignno{
{\delta \psi}_{\mu \nu} &= 
{\delta_0 \psi}_{\mu \nu} + \kappa \; {\delta_1 \psi}_{\mu \nu} 
\;\; , &(3.89a) \cr
{\delta_0 \psi}_{\mu \nu} &= 
- \Bigl[ \; 2\omega_{(\mu , \nu)} + 
2\eta_{\mu \nu} \; \omega^{\rho} \; (\ln \Omega)_{, \rho} \; \Bigr] 
\;\; , &(3.89b) \cr
{\delta_1 \psi}_{\mu \nu} &= 
- \Bigl[ \; 2\omega^{\rho}_{~, ( \mu} \; \psi_{\nu ) \rho} + 
\omega^{\rho} \; \psi_{\mu \nu , \rho} + 
2\omega^{\rho} \; \psi_{\mu \nu} \; (\ln \Omega)_{, \rho} \; \Bigr] 
\;\; , &(3.89c) }$$
where we have decomposed ${\delta \psi}_{\mu \nu}$ into a term with
no $\psi_{\mu \nu}$ and a term linear in $\psi_{\mu \nu}$. 
There is a similar transformation on the full scalar field:
$$\varphi'(x) = \varphi \Bigl( {y'}^{-1}(x) \Bigr) 
\;\; , \eqno(3.90)$$
and a corresponding infinitesimal variation on the fluctuating 
scalar field:
$$\eqalignno{
\delta\phi &= \delta_0\phi + \delta_1\phi
\;\; , &(3.91a) \cr
\delta_0\phi &= - \kappa \omega^0 \; {\widehat \phi}^{\;'}
\qquad ; \qquad
\delta_1\phi = - \kappa \omega^{\mu} \; \phi_{, \mu}
\;\; , &(3.91b) }$$

The ghost Lagrangian is taken to be:
$${\cal L}_{\rm gh} = 
- \Omega \; {\overline \omega}^{\mu} \; {\delta F}_{\mu} 
\;\; . \eqno(3.92)$$
As usual, the infinitesimal parameter of the coordinate transformation 
(3.87) is the anticommuting ghost field $\omega$ and ${\overline 
\omega}$ is the antighost field. Up to a surface term the ghost 
interaction Lagrangian is:
$$\eqalignno{ 
{\cal L}^{(3)}_{\rm gh} &=
-\Omega \; {\overline \omega}^{\mu} \; {\delta_1 F}_{\mu} &(3.93a) \cr
&= -\kappa \Omega^2 \; {\overline \omega}^{\mu , \nu} \; \Bigl[ \;
\psi_{\mu \rho} \; \omega^{\rho}_{~ , \nu} + 
\psi_{\nu \rho} \; \omega^{\rho}_{~ , \mu} + 
\psi_{\mu \nu , \rho} \; \omega^{\rho} +
2\psi_{\mu \nu} \; \omega^{\rho} \; (\ln \Omega)_{, \rho} 
\; \Bigr] \cr
&\;\;\;\; + \kappa 
{\Bigl( \Omega^2 \; {\overline \omega}^{\rho} \Bigr)}_{, \rho} \;
\Bigl[ \; \psi_{\mu \nu} \; \omega^{\mu , \nu} + 
\frac12 \psi_{, \nu} \; \omega^{\nu} +  
\psi \; \omega^{\nu} \; (\ln \Omega)_{, \nu} \; \Bigr] \cr
&\;\;\;\; + \kappa^2
\Omega^2 \; {\widehat \phi}^{\;'} \;
{\overline \omega}^{\mu} \; \eta_{\mu 0} \;
\phi_{, \rho} \; \omega^{\rho} 
\;\; , &(3.93b) }$$
and completely determines the ghost-graviton and ghost-scalar 
interactions.

The quadratic part of the ghost action is:
$$\eqalignno{
{\cal L}^{(2)}_{\rm gh} &= 
-\Omega \; {\overline \omega}^{\mu} \; {\delta_0 F}_{\mu} &(3.94a) \cr
&= {\overline \omega}^{\mu} \; \Bigl[ \;
{\rm D}_A \; {\overline \delta}_{\mu}^{~ \rho} + 
{\rm D}_B \; \delta_{\mu}^{~0} \; \delta_0^{~\rho} \; \Bigr] \;
\omega_{\rho} 
\;\; , &(3.94b) }$$
leading to the following ghost propagator:
$$i\Bigl[ {_{\mu}}\Delta_{\rho} \Bigr](x,x') = 
i\Delta_A(x,x') \; {\overline \eta}_{\mu \rho} -
i\Delta_B(x,x') \; \eta_{\mu 0} \; \eta_{\rho 0} 
\;\; . \eqno(3.95)$$
It is not possible to explicitly compute the ghost propagator for a
generic scale factor. However, in the case of the power law scale 
factors of Table 1 we get:
$$\eqalignno{
i\Bigl[ {_{\mu}}\Delta_{\rho} \Bigr]_{s=0}(x;x') &= \; 
i\Delta_{\frac12}(x,x') \; \Bigl[ \; 
{\overline \eta}_{\mu \rho} - \eta_{\mu 0} \; \eta_{\rho 0} \;
\Bigr] \cr
&= {1 \over 4\pi^2} \; 
{1 \over (x - x')^2} \;\;
\eta_{\mu \rho}
\;\; , &(3.96) }$$
$$\eqalignno{
i\Bigl[ {_{\mu}}\Delta_{\rho} \Bigr]_{s=\frac12}(x;x') &= \;
i\Delta_{\frac12}(x,x') \; {\overline \eta}_{\mu \rho} -
i\Delta_{\frac32}(x,x') \; \eta_{\mu 0} \; \eta_{\rho 0} \cr
&= {\eta_0^2 \over 8\pi^2 \eta^2 \; \eta^{'~2}} \;
\Biggl\{ \;
{2\eta\; \eta' \over (x - x')^2} \;\;
\eta_{\mu \rho}
+ \Bigl( \; 
\ln \Bigl[ k_0^2 (x - x')^2 \Bigr] \cr
&\qquad\qquad\qquad\qquad\qquad\qquad\quad\;\;
+ 2(\gamma - 1) \; \Bigr) \;
\eta_{\mu 0} \; \eta_{\rho 0}
\; \Biggr\} 
\;\; , &(3.97) }$$ 
$$\eqalignno{
i\Bigl[ {_{\mu}}\Delta_{\rho} \Bigr]_{s=\frac23}(x;x') &= \;
i\Delta_{\frac32}(x,x') \; {\overline \eta}_{\mu \rho} -
i\Delta_{\frac52}(x,x') \; \eta_{\mu 0} \; \eta_{\rho 0} \cr 
&= {\eta_0^4 \over 8\pi^2 \eta^3 \; \eta^{'~3}} \;
\Biggl\{ \;
{2\eta\; \eta' \over (x - x')^2} 
- \Bigl( \; 
\ln \Bigl[ k_0^2 (x - x')^2 \Bigr] 
+ 2(\gamma - 1) \; \Bigr) \;
\Biggr\} \; \eta_{\mu \rho} \cr
&- {\eta_0^4 \over 8\pi^2 \eta^4 \; \eta^{'~4}} \; 
\Biggl\{ \; 
\Bigl[ \; \frac32 (x - x')^2 
- 2\eta \; \eta' \; \Bigr] \; 
\Bigl( \; \ln \Bigl[ k_0^2 (x - x')^2 \Bigr]
+ 2\gamma \; \Bigr) \cr
&\qquad + 
4\eta \eta' + {9 \over k_0^2} 
- \frac{11}2 (\Delta x)^2 + \frac92 (\Delta\eta)^2
\; \Biggr\} \; \eta_{\mu 0} \; \eta_{\nu 0}
\;\; . &(3.98) }$$ 
$$\eqalignno{
i\Bigl[ {_{\mu}}\Delta_{\rho} \Bigr]_{s=+\infty}(x;x') &= \; 
i\Delta_{\frac32}(x,x') \; {\overline \eta}_{\mu \rho} - 
i\Delta_{\frac12}(x,x') \; \eta_{\mu 0} \; \eta_{\rho 0} \cr
&= {1 \over 8\pi^2 \eta_0^2} \;
\Biggl\{ \;
{2\eta\; \eta' \over (x - x')^2} \;\;
\eta_{\mu \rho}
- \Bigl( \; 
\ln \Bigl[ k_0^2 (x - x')^2 \Bigr] \cr
&\qquad\qquad\qquad\qquad\qquad\qquad\qquad
+ 2(\gamma - 1) \; \Bigr) \;
{\overline \eta}_{\mu \rho}
\; \Biggr\} 
\;\; , &(3.99) }$$ 
Again, when $s = +\infty$ , there is agreement with the previously 
obtained result in de Sitter spacetime [3].

\vfil\eject
\centerline{\bf 4. Corrections to the Newtonian gravitational force}

An elementary application of the perturbative tools developed above is 
the computation of the linearized response of the gravitational field 
to a point source. We may as well choose a matter dominated universe so 
that in the appropriate limit we can deduce the gravitational force law
and compare it with the Newtonian law we currently measure.

The gravitational force is obtained from the geodesic equation the
worldline $\chi^{\mu}(\tau)$ of a particle obeys:
$${\ddot \chi}^{\mu}(\tau) + 
\Gamma^{\mu}_{~\rho \sigma}[\chi(\tau)] \;
{\dot \chi}^{\rho}(\tau) \; {\dot \chi}^{\sigma}(\tau) = 0
\;\; . \eqno(4.1)$$
Any deviation of its worldline from the geodesic trajectory it would
occupy in the absence of gravitational sources will be attributed to
a gravitational force. To make contact with the Newtonian theory we 
shall consider the slow motion of the particle in a weak stationary 
gravitational field. In that case, the geodesic equation simplifies
considerably:
$${\ddot \chi}_i = {\kappa \over 2 \Omega^4} \; h_{00 , i}
\;\; , \eqno(4.2)$$
where we have chosen ${\dot \chi}^0(\tau) = \Omega^{-1}(\chi^0)$. 
The gravitational force is that part of the physical acceleration
which is not attributable to the Hubble flow. Hence we can write:
$$F_i = m \; \Omega \; {\ddot \chi}_i 
\;\; , \eqno(4.3)$$
which, with (4.2), gives:
$$F_i = {\kappa m \over 2 \Omega^3} \; h_{00 , i}
\;\; . \eqno(4.4)$$

The weak stationary gravitational field will be assumed as due to
the presence of a point source. The action of the source worldline 
$\chi_{\rm s}^{\mu}(\tau)$ is:
$${\cal S}_{\rm s} = - m_{\rm s} \int d\tau \;
{\sqrt {- g_{\mu \nu}[\chi_{\rm s}(\tau)] \; 
{\dot \chi}_{\rm s}^{\mu}(\tau) \; {\dot \chi}_{\rm s}^{\nu}(\tau)} } 
\;\; , \eqno(4.5)$$
and the resulting stress tensor:
$$T_{\rm s}^{\mu \nu}(x) = {m_{\rm s} \over \sqrt {-g(x)}} 
\int d\tau \;
{ { {\dot \chi}_{\rm s}^{\mu}(\tau) \; 
{\dot \chi}_{\rm s}^{\nu}(\tau) \;
\delta^{(4)} \Bigl( x - \chi_{\rm s}(\tau) \Bigr) }
\over
{\sqrt {- g_{\rho \sigma}[\chi_{\rm s}(\tau)] \; 
{\dot \chi}_{\rm s}^{\rho}(\tau) \; 
{\dot \chi}_{\rm s}^{\sigma}(\tau)} } }
\;\; . \eqno(4.6)$$
The stationary nature of the source allows us to choose:
$${\dot \chi}_{\rm s}^{\mu}(\tau) = 
\Omega^{-1}(\chi^0) \; \delta^{\mu}_{~0}
\; \; , \eqno(4.7)$$
in which case the stress tensor simplifies to:
$$T_{\rm s}^{\mu \nu}(\eta, {\vec x}) = 
m_{\rm s} \; \Omega^{-5}(\eta) \;
\delta^{\mu}_{~0} \; \delta^{\nu}_{~0} \; \delta^{(3)}({\vec x})
\;\; . \eqno(4.8)$$

The linearized response of the gravitational field to the presence 
of the point source is given by [3]:
$$\psi_{\mu \nu}(\eta, {\vec x}) = 
- {\kappa \over 2} \int d\eta' \; \int d^3x' \; \Omega^6(\eta') \;
\Bigl[ {_{\mu\nu}}{G^{\rm ret}}_{\rho\sigma} \Bigr](x;x') \;
T_{\rm s}^{\rho \sigma}(x')
\;\; , \eqno(4.9)$$
where the pseudo-graviton retarded Green's function is defined as:
$$\Bigl[ {_{\mu\nu}}{G^{\rm ret}}_{\rho\sigma} \Bigr](x;x') 
\equiv
2 \theta(\Delta \eta) \;\; 
{\rm Im} \Biggl[ \;
i\Bigl[ {_{\mu \nu}}\Delta_{\rho \sigma} \Bigr](x;x') 
\; \Biggr]
\;\; . \eqno(4.10)$$ 
For the power law scale factors we have analyzed, the retarded 
propagators $G_I^{\rm ret}(x;x')$ can be expressed as follows:
$$G_I^{\rm ret}(x;x') = 2 \theta(\Delta \eta) \;\; 
{\rm Im} \Bigl[ \; i\Delta_I^s(x;x') \; \Bigr]
\;\; . \eqno(4.11)$$
In terms of the mode functions: 
$$G_I^{\rm ret}(x;x') = {\theta(\Delta\eta) \over \pi^2 \; \Delta x}
\int_0^{+\infty} dk \; 
k \; {\rm sin}(k \Delta x) \; 
{\rm Im} \Bigl[ \; 
\Psi(\eta, k, I) \;\; \Psi^*(\eta', k, I) \; \Bigr]
\;\; , \eqno(4.12)$$
where we have used (3.43b).

Inspection of the form of the three tensor factors (3.41) and the
stress tensor (4.8) shows that only the $\delta_{\mu}^{~0} \;
\delta_{\nu}^{~0} \; \delta_{\rho}^{~0} \; \delta_{\sigma}^{~0}$ 
part of $\Bigl[ {_{\mu\nu}}{T^C}_{\rho\sigma} \Bigr]$ can contribute
to the desired $\kappa \psi_{00}(\eta, {\vec x})$ response.
$$\psi_{00}(\eta, {\vec x}) = 
-{\kappa \over 2} \int d\eta' \; \int d^3x' \; \Omega^6(\eta') \;
\Bigl[ \frac1{1+s} \; G_A^{\rm ret}(x;x') +
\frac{s}{1+s} \; G_C^{\rm ret}(x;x') \Bigr] \;
\delta_{\rho}^{~0} \; \delta_{\sigma}^{~0} \; 
T_{\rm s}^{\rho \sigma}(x')
\;\; . \eqno(4.13)$$
Specializing to the matter dominated universe leads to:
$$\eqalignno{
G_A^{\rm ret} \; \Big\vert_{s = \frac23} \; (x;x') &= 
{\theta(\Delta\eta) \over 2\pi^2 \; \Delta x} \;
{\eta_0^4 \over \eta^3 \; {\eta'}^3} 
\int_0^{+\infty} dk \; {\rm sin}(k \Delta x) \;
\Biggl\{ \;
{\rm cos}(k \Delta\eta) \; \Bigl[ \;
{\Delta\eta \over k} \; \Bigr] 
&(4.14a) \cr
&\qquad\qquad\qquad\qquad\qquad\qquad\quad
+ {\rm sin}(k \Delta\eta) \; \Bigl[ \;
-{1 \over k^2} - \eta \; \eta' \; \Bigr] \;
\Biggr\}
\;\; , \qquad\qquad }$$
$$\eqalignno{
G_C^{\rm ret} \; &\Big\vert_{s = \frac23} \; (x;x') = 
{\theta(\Delta\eta) \over 2\pi^2 \; \Delta x} \;
{\eta_0^4 \over \eta^5 \; {\eta'}^5} 
\int_0^{+\infty} dk \; {\rm sin}(k \Delta x) \;\; \times
&(4.14b) \cr
&\qquad\qquad\qquad\quad
\Biggl\{ \;
{\rm cos}(k \Delta\eta) \; \Bigl[ \;
{225 \Delta\eta \over k^5} 
- {15 (\Delta\eta)^3 \over k^3}
+ {45 \eta \; \eta' \; \Delta\eta \over k^3} 
+ {6 \eta^2 \; {\eta'}^2 \; \Delta\eta \over k} 
\; \Bigr] \cr
&\;
+ {\rm sin}(k \Delta\eta) \; \Bigl[ \;
- {225 \over k^6} 
+ {90 (\Delta\eta)^2 \over k^4}
- {45 \eta \; \eta' \over k^4}
+ {15 \eta \; \eta' \; (\Delta\eta)^2 \over k^2}
- {6 \eta^2 \; {\eta'}^2 \over k^2} 
- \eta^3 \; {\eta'}^3 \; \Bigr] 
\; \Biggr\}
\;\; , }$$
where we have used Table~4 to substitute the appropriate functional 
form of the mode function and its complex conjugate in (4.12). We 
then perform all integrations by parts needed to express (4.14) in
terms of the following integrals:
$$\eqalignno{
I_{\rm sin} \equiv \int_0^{+\infty} dk \; 
{\rm sin}(k \Delta x) \; {\rm sin}(k \Delta\eta) &=
\frac{\pi}2 \delta (\Delta x - \Delta\eta)
- \frac{\pi}2 \delta (\Delta x + \Delta\eta)
\;\; , &(4.15a) \cr
I_{\rm cos} \equiv \int_0^{+\infty} {dk \over k} \; 
{\rm cos}(k \Delta x) \; {\rm sin}(k \Delta\eta) &=
\frac{\pi}2 \theta (\Delta x + \Delta\eta)
- \frac{\pi}2 \theta (\Delta x - \Delta\eta)
\;\; . &(4.15b) }$$
The result is:
$$\eqalignno{
G_A^{\rm ret} \; \Big\vert_{s = \frac23} \; (x;x') &= 
- {\theta(\Delta\eta) \over 2\pi^2} \; 
{\eta_0^4 \over \eta^2 \; {\eta'}^2} \;
\Biggl\{ \;
{1 \over \Delta x} \; I_{\rm sin}
+ {1 \over \eta \; \eta'} \; I_{\rm cos}
\; \Biggr\}
\;\; , &(4.16a) \cr
G_C^{\rm ret} \; \Big\vert_{s = \frac23} \; (x;x') &= 
- {\theta(\Delta\eta) \over 2\pi^2} \; 
{\eta_0^4 \over \eta^2 \; {\eta'}^2} \;
\Biggl\{ \;
{1 \over \Delta x} \; I_{\rm sin}
+ \Bigl[ \; 
{6 \over \eta \; \eta'}
- {15(x -x')^2 \over 2\eta^2 \; {\eta'}^2} \cr
&\qquad\qquad\qquad\qquad\qquad\qquad\qquad\quad
+ {15(x - x')^4 \over 8\eta^3 \; {\eta'}^3}
\; \Bigr] \; I_{\rm cos}
\; \Biggr\} 
\;\; , \qquad &(4.16b) }$$
which finally implies:
\footnote{*}{\tenpoint These retarded Green's functions could 
also be obtained directly from the imaginary part of the 
$i\Delta_{\frac72}(x;x')$ propagator (3.83).} 
$$\eqalignno{
G_A^{\rm ret} \; \Big\vert_{s = \frac23} \; (x;x') &= 
- {\theta(\Delta\eta) \over 4\pi} \;
{\eta_0^4 \over \eta^2 \; {\eta'}^2} \;
\Biggl\{ \;
{1 \over \Delta x} \; \delta(\Delta x - \Delta\eta) 
+ {1 \over \eta \; \eta'} \; \theta(\Delta\eta - \Delta x)
\; \Biggr\} 
\;\; , &(4.17a) \cr
G_C^{\rm ret} \; \Big\vert_{s = \frac23} \; (x;x') &= 
- {\theta(\Delta\eta) \over 4\pi} \; 
{\eta_0^4 \over \eta^2 \; {\eta'}^2} \;
\Biggl\{ \;
{1 \over \Delta x} \; \delta(\Delta x - \Delta\eta) 
+ \Bigl[ \; 
{6 \over \eta \; \eta'}
- {15(x -x')^2 \over 2\eta^2 \; {\eta'}^2} \cr
&\qquad\qquad\qquad\qquad\qquad\qquad\qquad
+ {15(x - x')^4 \over 8\eta^3 \; {\eta'}^3}
\; \Bigr] \; \theta(\Delta\eta - \Delta x)
\; \Biggr\} 
\;\; . \qquad\qquad &(4.17b) }$$

When we insert (4.8) and (4.17) in equation (4.13) and use the 
delta function present in $T_{\rm s}^{\rho \sigma}(x')$ to do 
all three spatial integrations, we get:
$$\psi_{00}(\eta, {\vec x}) = 
{\kappa \; m_{\rm s} \over 8\pi} \; 
{\eta_0^2 \over \eta^2} \;
\Biggl\{ \;
{1 \over x} \; + \;
\int_{\eta_1}^{\eta - x} d\eta' 
\Bigl[ \;
{3(x^2 - \eta^2)^2 \over 4\eta^3 \; {\eta'}^3}
\; - \; {3(x^2 - \eta^2) \over 2\eta^3 \; \eta'}
\; + \; {3\eta' \over 4\eta^3}
\; \Bigr] 
\; \Biggr\} 
\;\; , \eqno(4.18)$$
where $x = \Vert {\vec x} \Vert$ and $\eta_1$ is the beginning 
of matter domination in the universe. In terms of the graviton
field $h_{\mu \nu} = \Omega^2 \psi_{\mu \nu}$ the conformal time
integration gives:
$$h_{00}(\eta, {\vec x}) = 
{\kappa \; m_{\rm s} \over 16\pi} \; \Omega \;
\Biggl\{ \;
{2 \over x} \; - \;
{3\eta_1^2 \over 4\eta^3} \; - \; 
{3x \over \eta^2} \; - \;
{3(x^2 -\eta^2) \over \eta^3} \; 
\ln \Bigl( {\eta - x \over \eta_1} \Bigr) \; + \;
{3(x^2 - \eta^2)^2 \over 4\eta^3 \; \eta_1^2}
\; \Biggr\} 
\;\; . \eqno(4.19)$$

The gravitational force is obtained from (4.4):
$$\eqalignno{
F(\eta, r) = - G \; m \; m_{\rm s} \;
\Biggl\{ \;
{1 \over r^2} \; + \;
{3 \over \Omega^2 \; \eta^2} \; + \;
{3r \over 2\Omega^3 \; \eta^3} \; + \;
{3r \over 2\Omega^3 \; \eta \; \eta_1^2} 
\; &+ \;
{3r \over \Omega^3 \; \eta^3} \; \ln \Bigl( 
{\Omega \eta - r \over \Omega \eta_1} \Bigr) \cr
&- \; 
{3r^3 \over 2\Omega^5 \; \eta^3 \; \eta_1^2}
\; \Biggr\}
\;\; , \qquad\qquad &(4.20) }$$
and has been expressed in terms of the physical distance ${\vec r} 
= \Omega \; {\vec x}$. The first term in (4.20) is the Newtonian 
attractive inverse square law. The remaining terms -- one of which 
is independent of the distance -- represent the corrections. To get 
an estimate for their relative strength, we apply (1.3)-(1.4) to a 
matter dominated universe: 
$$\eta = 3 \; t_0^{\frac23} \; t^{\frac13}
\qquad ; \qquad
\eta_0 = 3 \; t_0 
\;\; , \eqno(4.21)$$
and re-express the force as:
$$\eqalignno{
F(t, r) = - G \; m \; m_{\rm s} \;
\Biggl\{ \;
{1 \over r^2} \; + \;
{1 \over 3t^2} \; + \;
{r \over 18t^3} \; + \;
\frac1{18} \; r \; t^{-\frac73} \; t_1^{-\frac23}
\; &+ \;
{r \over 9t^3} \; \ln \Bigl[ 
(t - \frac13 r) \;
t^{-\frac23} \; t_1^{-\frac13} \Bigr] \cr
&- \; 
\frac1{162} \; r^3 \; t^{-\frac{13}3} \; t_1^{-\frac23} 
\; \Biggr\}
\;\; , \qquad\qquad &(4.22) }$$
The beginning of matter domination is thought to have occured
at a physical time $t_1$ such that [11]:
$${a(t) \over a(t_1)} \approx 10^4
\qquad \Longrightarrow \qquad
t = 10^6 \; t_1
\;\; , \eqno(4.23)$$
where, of course, $t$ is the present. Using (4.23) we obtain:
$$F(t, r) = - {G \; m \; m_{\rm s} \over r^2} \;
\Biggl\{ \; 1 \; + \;
\frac13 \; \Bigl( {r \over t} \Bigr)^2 \; + \;
\frac{5000}9 \; \Bigl( {r \over t} \Bigr)^3 \; + \;
\frac19 \; \Bigl( {r \over t} \Bigr)^3 \; \ln \Bigl[
\frac{100}3 \Bigl( 3 - {r \over t} \Bigr) \Bigr] \; - \;
\frac{5000}{81} \; \Bigl( {r \over t} \Bigr)^5
\; \Biggr\}
\eqno(4.24)$$
It is now apparent that the correction terms are negligible unless
we consider objects of size comparable to that of the observed
universe. The present time $t$ is of the order of $10^{28}{\rm cm}$ while
a galaxy and a cluster of galaxies have typical size $r \approx 
10^{23}{\rm cm}$ and $r \approx 10^{25}{\rm cm}$ respectively.

\vskip 1cm
\centerline{\bf 5. Epilogue}

One possible explanation of the discrepancy between the predictions
of the Newtonian theory and the data provided by the galactic rotation 
curves, is an appropriate modification of the Newtonian gravitational 
force law at large distances. A natural way for this to occur would be
due to the deviations from flat space introduced by the matter dominated
background. These deviations turn out to be unable to bridge the
discrepancy by themselves or to, at least, reduce significantly the 
large amount of dark matter required for that purpose.

In retrospect, it would have been very hard for such corrections to
have the desired effect. For if they were strong at galactic scales to
resolve the rotation curves mystery, they would become enormous at even
larger scales something which is ruled out by the existence of structures
much larger than galaxies: there is no gravitational confinement at
galactic scales. The corrections would have to be of the proper strength
at galactic scales and weak at all other scales.

From the quantum field theoretic point of view, the Feynman rules were
developed for a spatially flat Robertson-Walker background generated by
a dynamical scalar source. This is a much wider class of backgrounds
than what is needed for the currently observed universe. Their study
revealed that conformal flatness is a more powerful organizing principle
than maximal symmetry. This fact becomes especially obvious from the
identity of the expressions for the quadratic Lagrangian in the case of
de Sitter spacetime (maximally symmetric) and flat Robertson-Walker
spacetimes (not maximally symmetric). 

Moreover, the Feynman rules obtained make possible any perturbative 
calculation desired for power law scale factors. For instance, the 
existence of strong infrared quantum gravitational effects could be 
studied in detail in these backgrounds.

\vskip 1cm
\centerline{ACKNOWLEDGEMENTS}

This work was partially supported by DOE contract DE-FG02-97ER41029, 
by NSF grant 94092715, by EU grants CHRX-CT94-0621 and TMR96-1206, 
by ${\rm \Gamma \Gamma ET}$ grant \break
${\rm \Pi ENE \Delta}$-95-1759, by 
NATO grant CRG-971166, and by the Institute for Fundamental Theory.

\vfil\eject

\centerline{REFERENCES}

\item{[1]} H. P. Robertson, {\sl Ap. J.} {\bf 82} (1935) 284 ;
{\bf 83} (1936) 187 ; {\bf 83} (1936) 257.
\hfill\break  
A. G. Walker, {\sl Proc. Lond. Math. Soc. (2)} 
{\bf 42} (1936) 90.

\item{[2]} C. L. Bennett et al., {\sl Astrophys. J. Lett.} 
{\bf 464} (1996) L1.

\item{[3]} N. C. Tsamis and R. P. Woodard, {\sl Commun. Math. Phys.} 
{\bf 162} (1994) 217.

\item{[4]} N. D. Birrell and P. C. W. Davies, {\it Quantum Fields in 
Curved Space} (Cambridge University Press, Cambridge, UK, 1982).

\item{[5]} N. C. Tsamis and R. P. Woodard, {\sl Phys. Lett.} 
{\bf 292B} (1992) 269.

\item{[6]} B. Allen and M. Turyn {\sl Nucl. Phys.} 
{\bf 292B} (1987) 213.
\hfill\break  
I. Antoniadis and E. Mottola, {\sl J. Math. Phys.} 
{\bf 32} (1991) 1037.

\item{[7]} E. G. Floratos, J. Iliopoulos and T. N. Tomaras, 
{\sl Phys. Lett.} {\bf 197B} (1987) 373.

\item{[8]} N. C. Tsamis and R. P. Woodard, {\sl Class. Quantum Grav.} 
{\bf 11} (1994) 2969.

\item{[9]} L. H. Ford and L. Parker, {\sl Phys. Rev.} {\bf D16} (1977) 245;
\hfill\break
J. Iliopoulos and T. N. Tomaras, {\sl Phys. Lett.} {\bf 255B} (1991) 27; 
{\sl erratum} {\bf 263B} (1991) 591.

\item{[10]} I. S. Gradshteyn and I. M. Ryzhik, {\it Table of Integrals, 
Series, and Products} (Academic Press, New York, USA, 1965).

\item{[11]} E. W. Kolb and M. S. Turner, {\it The Early Universe} 
(Addison-Wesley, Redwood City, USA, 1990).

\bye